**Adhesion of fluid infused silicone elastomer to glass**

Anushka Jha[1], Preetika Karnal[1,2], and Joelle Frechette[1,3,*]

*1. Chemical and Biomolecular Engineering Department, Johns Hopkins University, Baltimore, MD 21218, USA 2. Department of Chemical and Biomolecular Engineering, Lehigh University, 124 E Morton St, Building 205, Bethlehem, Pennsylvania 18015, United States 3. Chemical and Biomolecular Engineering Department, University of California, Berkeley, CA 94760, USA*

*Corresponding author: Joelle Frechette, email: jfrechette@berkeley.edu

**ABSTRACT**

Elastomers swollen with non-polar fluids show potential as anti-adhesive materials. We study the effect of oil fraction and contact time on the adhesion between swollen spherical probes of PDMS (polydimethylsiloxane) and flat glass surfaces. The PDMS probes are swollen with pre-determined amount of 10 cSt silicone oil to span the range where the PDMS is fluid free (via solvent extraction) up to the limit where it is oil saturated. Probe tack measurements show that adhesion decreases rapidly with an increase in oil fraction. The decrease in adhesion is attributed to excess oil present at the PDMS-air interface. Contact angle measurements and optical microscopy images support this observation. Adhesion also increases with contact time for a given oil fraction. The increase in adhesion with contact time can be interpreted through different competing mechanisms that depend on the oil fraction where the dominant mechanism changes from extracted to fully swollen PDMS. For partially swollen PDMS, we observe that adhesion initially increases because of viscoelastic relaxation and at long times increases because of contact aging. In contrast, adhesion between fully swollen PDMS and glass barely increases over time and is mainly due to capillary forces. While the relaxation of PDMS in contact is well-described by a visco-poroelastic model, we do not see evidence that poroelastic relaxation of the PDMS contributes to an increase of adhesion with glass whether it is partially or fully swollen.

# INTRODUCTION

Oil infused elastomers are crosslinked polymer networks that have undergone swelling by a nonpolar fluid. They have shown promise as low friction and anti-adhesive materials.[1, 2] In particular, silicone elastomers such as PDMS (polydimethylsiloxane), swollen by linear silicone oils result in slippery materials due to the presence of excess oil on their surface. Silicone oil can be added to the network by immersing the elastomer in the oil after curing[3, 4], be added during curing[5], or be present as unreacted oligomers[6]. Oil infused elastomers have found applications toward anti-icing[5, 7-13], or antifouling.[4, 14-16] Moreover, the presence of fluid within elastomers has enabled their use as skin mimics[17-22] or light responsive materials.[20, 23] To develop and optimize oil-infused elastomers for these technologies, a better understanding of how the presence of oil within the matrix alters their surface and adhesive properties is needed.

A large body of work on adhesion and surface properties of swollen elastomers is in the development of anti-icing surfaces. Ice adhesion (commonly measured in shear) decreases when oils are incorporated within the PDMS matrix, and the decrease is attributed to the presence of oil at the solid/air interface.[5, 24] Contact angle measurements using oil droplets shows that the area fraction of oil on swollen elastomers increases linearly with the amount of oil in the bulk up to complete coverage.[5] For fully swollen PDMS, contact angle measurements show the surface to have a low contact angle hysteresis.[3, 4, 9, 25, 26] Optical microscopy also confirms the presence of oil at the PDMS/air interface.[3, 7, 8] Beyond ice adhesion, studies of adhesion to swollen (or partially swollen) PDMS are limited. In particular, the relationship between adhesive strength and oil fraction for contact with surfaces other than ice has yet to be investigated.

The presence of oil within an elastomer matrix alters its surface and bulk properties, both determinant of adhesive strength. Oil within the bulk of the elastomer will also be present at the PDMS/air interface, altering the surface energy. In addition, even small amounts of oil present can significantly affect the wetting dynamics of a PDMS surface.[26] Experiments also show that if the surface oil is physically removed, by wiping for example, it is quickly replenished due to the lower surface energy of the fluid compared to the elastomer.[3, 4, 9] In addition, when the surface and bulk elastic stresses are comparable in an elastomer, elasto-capillarity also plays a significant role in determining the surface deformation and wetting of fluid-infused elastomers.[27-30] We expect surface energy, oil replenishment, and elasto-capillarity to be affected by the oil fraction and contribute to adhesion.

Silicone oil within the matrix also alters the bulk mechanical response of an elastomer.[3, 9] In the absence of fluid, the viscoelastic portion of the stress response of PDMS is described using a generalized Maxwell model.[31] When fluid is present in the bulk, elastomer relaxation is no longer solely a function of the elastomer network (viscoelasticity) but also depends on fluid transport through the pores – also known as poroelasticity.[31-36] The poroelastic response is typically described as a single exponential with $\tau_P = a^2/D$ as the timescale for exponential decay where $a$ is a length-scale describing size of contact and $D$ is the effective diffusivity.[33, 37] Models describing the dynamic stress response of fully swollen elastomers consider both viscoelasticity and poroelasticity – often describing the net response as the sum of the two.[33, 38, 39] In addition, experiments have shown that upon deformation of the swollen elastomer, fluid from the bulk can get transferred to the interfacial region due to local stress gradients.[27, 28, 30] The oil pushed to the interface creates a four phase contact zone at the periphery of an indenter where the volume of fluid pushed out depends on indentation depth and compressibility.[30] How this dynamic relaxation processes affect adhesion for swollen PDMS remains to be investigated. Moreover, decoupling the relative contributions of interfacial effects (solid-solid contact, capillarity) from bulk contributions (poroelasticity and viscoelasticity) on adhesion to swollen elastomers could help expand and tune their properties for ice adhesion and other applications.

The measurement of stress response and adhesion as a function of dwell time has been instrumental in understanding the dynamics of adhesion for extracted (dry) PDMS.[40-42] Similarly, experiments with hydrogels suggest that multiple dynamic phenomena such as poroelasticity, viscoelasticity, and muco-adhesion contribute to adhesive properties of fluid filled networks. For example, such experiments have been instrumental as models of fluid filled networks for understanding relaxation processes[34, 35, 43, 44] and adhesion[32, 45, 46] (and sometimes, the relationship between the two). Hydrogels generally contain more fluid in their network than elastomers, they routinely include $> 60 - 90\%$ water. Due to the high water content, most hydrogel relaxation studies are conducted in submerged conditions.[45-52] Peeling measurements have been done in air with hydrogels with high fluid content or in presence of another saturated surface.[53] For hydrogels, poroelastic relaxation leads to an increase in adhesion for contact times that are longer than the poroelastic time scale (determined based on the effective diffusivity of water within the network and the contact radius). The increase in adhesion is attributed to a "suction" pressure that develops across the interface due to an osmotic pressure gradient.[45, 46] Experiments with hydrogels also show that adhesion increases with contact time if the fluid can be transferred to the opposing surface, a mechanism known as muco-adhesion.[46, 53] Curatolo et. al. conducted loop test experiments and found that muco-adhesion was also present in the contact between elastomers, where dynamic swelling of both surfaces lead to an increase in surface energy and adhesive strength.[54]

Here, we characterize the normal interactions between a PDMS probe and a glass surface (**Fig. 1**). In addition to adhesion measurements, we also characterize the surface properties and bulk relaxation behavior of the elastomers. Specifically, we hypothesize that the fraction of the oil within the elastomer and contact time will both influence the adhesive properties of swollen elastomers. Because the silicone oils employed are non-volatile (no fluid loss from evaporation), we can explore the regime where the network is swollen with oil in air. We observe a transition from solid-solid adhesion to capillary adhesion as the oil fraction increases. We also conduct *in-situ* relaxation followed by adhesion to measure adhesion at varying degrees of relaxation. We demonstrate the importance of viscoelasticity and solid-solid contact formation at the interface on adhesion for swollen elastomers. We will also see that poroelasticity is observed during relaxation of these elastomers, but it does not contribute to adhesion.

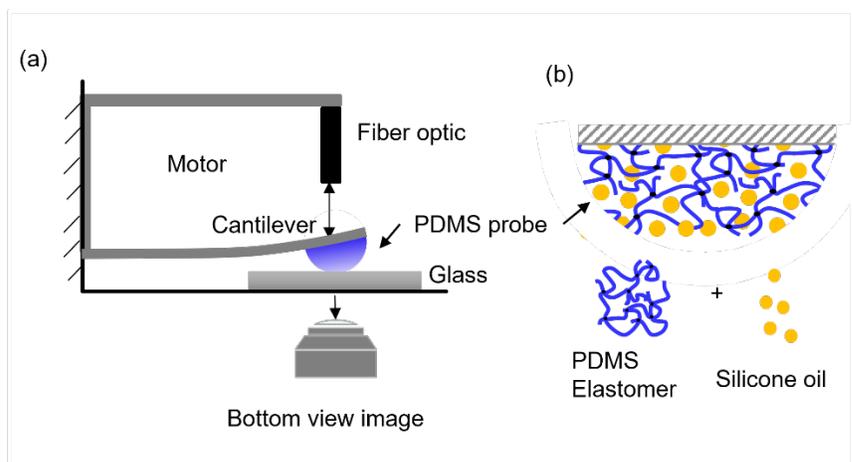

**Fig. 1. Schematic of set-up to study normal interactions between PDMS probes and glass surfaces.** (a) Experimental configuration to measure adhesion, (b) Enlarged schematic of PDMS lens swollen in silicone oil used for adhesion measurements.

## MATERIALS AND METHODS

Materials. We use Sylgard$^{TM}$ 184, a two-part silicone elastomer kit, for the PDMS (Fisher Scientific). Silicone oil (polydimethylsiloxane, trimethylsiloxy terminated) of viscosity 10 cSt (Gelest) was used as received and sealed with Parafilm to prevent contamination. All the solvents, acids and bases used were ACS grade (Sigma Aldrich).

Preparation of PDMS probes and swelling with silicone oil. Hemispherical PDMS lenses are fabricated by first curing a flat, 4 mm thick PDMS sheet to serve as a base, followed by drop casting a drop on top of the cured PDMS sheet, followed by a second curing. For both steps a 10:1 mixture of Sylgard 184 (10-parts prepolymer, 1-part crosslinker) is mixed in a petri dish and degassed in a vacuum chamber to remove air-bubbles. The mixture is then cured in the petri dish at 75 °C, 1 atm for ~16 hours. After curing, a 6 mm circle is punched out of the sheet to be used as the base of the hemispherical lens. Then, 40 $\mu L$ of uncured PDMS mixture is drop cast onto the base.[55] Pinning of the mixture at the edge of the base leads to a spherical cap that is then cured again. The samples are then soaked in n-hexane for a duration of 6 hours to extract unreacted oligomers remaining after the curing process. The percentage weight loss in the sample becomes constant after 5 hours of extraction.[56] Following extraction, we remove the hexane from the elastomer by soaking the samples in an ethanol bath for 20 minutes in a sonicator, followed by drying at 75°C, 1 atm overnight. The amount of oligomer extracted is $3.6 \pm 0.1\ wt\%$ using our method, which is comparable with the $4.5 \pm 0.9\%$ obtained by Glover *et. al.* after 24 h hexane extraction.[57] The final samples are hemispherical lenses with radius of curvatures of approximately 6 mm (**Fig. S1**, ESI†) and weight ~0.1 g.

We create swelling curves and by determining the oil fraction as a function time submerged for 10 cSt silicone oil in PDMS. To do so, the PDMS lenses are soaked in 10 cSt silicone oil bath at room temperature for a set amount of time. Upon removal from the bath, we use pressurized nitrogen gas to remove excess oil from the surface. The samples are then weighed, and the time elapsed during the swelling process is recorded to obtain a swelling curve (**Fig. S2**, ESI†). We calculate the oil fraction $\phi$ by measuring the initial and final mass of the swollen elastomer, $m_i$ and $m_f$, respectively as given by **Eq. (1).**

$$\phi = \frac{m_f - m_i}{m_f} \tag{1}$$

Because the density of the silicone oil and the elastomer are very similar (~935 kg/m³), the mass oil fraction $\phi$ is nearly identical to the volume oil fraction ($\Lambda$). The lens takes approximately 4 days to be fully saturated with oil, with a mass oil fraction of $\phi_{sat} = 0.40 \pm 0.03$ and volume oil fraction $\Lambda_{swell} = \frac{V_{sat} - V_{dry}}{V_{dry}} = 0.44 \pm 0.01$.

From the swelling curve, we can immerse PDMS lenses in silicone oil for preset amounts of time to achieve a predetermined oil fraction ranging from $0 < \phi \leq 0.4$. After swelling, the lenses are glued on top of a 9mm optical window (Edmund optics) using a thin layer(<5 μL) of uncured 10:1 Sylgard mixture. The lens-window system is placed on a hotplate at 150 °C for ~15 s to solidify the thin layer of glue. Each lens-window pair is used only for one contact experiment and discarded afterwards.

<u>Glass substrate preparation.</u> Glass microscope slides (46 mm x 27 mm, Ted Pella Inc) are sonicated first in ethanol, followed by isopropyl alcohol each for 20 minutes, and then cleaned with piranha solution (3:1, sulfuric acid: hydrogen peroxide) for 1 hr. The glass slides are then stored in deionized (DI) water and used within a day for adhesion measurements. Just prior to the measurements the slides are dried using pressurized nitrogen. The cleaned glass slides are hydrophilic in nature with a water contact angle $\theta_{water\ on\ glass} < 10°$.

<u>Rheology.</u> Rheological measurements were performed on 1 mm thick discs of cured elastomer on an Anton Paar MCR 302 using an 8 mm parallel plate accessory. The measurements were conducted for different oil fractions. Frequency sweep (1 rad/s to 100 rad/s) measurements were carried out at 1% strain under a normal force of 10 N. Storage ($G'$) and loss ($G''$) moduli were recorded as a function of angular frequency $\omega$ for different oil fraction (see **Fig. S3**, ESI†).The storage modulus changes slightly: from $586 \pm 44$ kPa to $630 \pm 67$ kPa at $f \sim \omega/2\pi \sim 1$Hz, for different oil fraction. The loss modulus decreases sharply from $58 \pm 1$ kPa to $5 \pm 1$ kPa with increasing oil fraction. The storage modulus is about two orders of magnitude higher than the loss modulus and $G''/G' = \tan\delta \ll 1$ for all oil fractions implying that the material is highly elastic. The shear modulus $G = \sqrt{G'^2 + G''^2}$ changes only slightly from $589 \pm 43$ kPa to $630 \pm 67$ kPa with increasing oil fraction. The Young's modulus of dry PDMS, $E = 2G(1+\nu)$ is $2.1 \pm 0.3$ MPa (for (Poisson ratio $\nu \approx 0.5$), consistent with literature reports on mechanical properties of Sylgard 184.[57-59] Additionally, the equilibrium volume oil fraction obtained from swelling experiments, $\Lambda_{swell} = 0.4$ is found to be comparable with predictions from Flory-Rehner theory ($\Lambda_{F-R} = 0.42$) for the fully-swollen elastomer[60].

Contact angle measurements and optical microscopy. Contact angles were obtained from sessile droplets using a goniometer (Dataphysics OCA, **Fig. S4**, ESI†). The surface energy was calculated from the two liquid method, using DI water and Diiodomethane with $5\mu L$ drops on the PDMS substrate.[61] Diiodomethane and DI water have very low solubility[62, 63] in PDMS and are therefore used as ideal apolar and polar liquids respectively to measure the surface energy of PDMS. [64] We also imaged the PDMS surface using an optical microscope in brightfield (NI instruments).

Poroelastic Relaxation Indentation and probe tack measurements. Poroelastic relaxation indentation (PRI) measurements[31] were carried out in a homebuilt multifunctional force microscope (MFM) with bottom view imaging (**Fig. S5**, ESI†).[60, 65] The PDMS lens is mounted on a cantilever with spring constant $k_{spring}$. The lens is slowly lowered ($v \sim 5~\mu m/s$) onto the glass surface while the approach is monitored using a side view camera. The motor is stopped as soon as Newton's rings become visible in the bottom view image (**Fig. S6**, ESI†). At this point, the approach velocity is increased to $v = 500~\mu m/s$ to induce step-indentation and the cantilever is lowered by a set distance of $\Delta M = 50~\mu m$. A fiber optic system measures the spring deformation $\delta_{spring}$ and displays the spring force $F = k_{spring} \times \delta_{spring} \sim 60~mN$. The indentation depth during each experiment is calculated using $\delta = \Delta M - \delta_{spring} \approx 35~\mu m$.

The two surfaces are kept in contact for varying dwell times $\tau_D = 10s - 10800~s$. Although the indentation depth slightly increases over time due to relaxation of the elastomer, it can be assumed to be constant since the change is negligible compared to the magnitude of indentation depth $\frac{\Delta\delta}{\delta} \sim 1\%$. After dwell, the cantilever is then retracted at $v = -50~\mu m/s$ and the debonding force and contact area are recorded as a function of time.

Probe-tack experiments are conducted in the sphere-plane geometry. The PDMS lens approaches the glass surface at $v = 50~\mu m/s$ until it reaches a constant dwell force of 10 mN. The dwell force is maintained at 10 mN via force feedback control system for varying dwell time and the cantilever is retracted at $v = -50\ \mu m/s$. The approach velocity is lower during probe-tack in order to reach a stable set-point of dwell force within O (0.1s). The force and contact area are recorded over time. The experimental parameter for the PRI and probe tack measurements are given in **Table 1**.

**Table 1.** Experimental parameters for probe-tack and PRI.

| Parameter | | Probe tack | PRI |
|---|---|---|---|
| Spring constant | $k_{spring}$ | 4499 N/m | 1091 N/m |
| Approach velocity | $v_{approach}$ | 50 μm/s | 500 μm/s |
| Dwell condition | | Fixed load, 10 mN | Fixed indentation depth, $\delta \sim 35$ μm |
| Retraction velocity | $v_{debond}$ | -50 μm/s | -50 μm/s |

<u>Visco-poroelastic modeling of PRI data.</u> The force vs time data spans four decades (10 s -$10^4$ s) in time and we sample points on a logarithmic scale. The data recorded during relaxation experiments is fitted to visco-poroelastic model using non-linear least squares fitting.[31] The Gauss-Newton method is implemented in Python 3 and a tolerance of $10^{-8}$ is set on the cost function. To ensure stable convergence, we also monitor the values after each iteration. The goodness of the fit is determined for each fit by the standard error which is always less than 10%.

## RESULTS AND DISCUSSION

### A. Characteristic elastomer timescales

We begin by identifying the characteristic relaxation timescales of the elastomer for different oil fractions. To do so, we conduct PRI measurements where we measure the force relaxation at (near) constant indentation depth of ~35 $\mu m$ as a function of time for over ~$10^4\ s$ (**Fig. 2**). We compare the relaxation of PDMS for three different oil fractions: $\phi = 0, 0.1,$ and $\phi = 0.4$ in air as well as $\phi = 0.4$ in oil. (Note that $\phi = 0.4$ corresponds to saturation). For all $\phi$ investigated the final force reaches a comparable value at long time, but the relaxation curves and initial forces exhibit distinct features.

We describe the relaxation curves using a visco-poroelastic model **Eq. (2)** based on the Maxwell-Wiechert model used previously by Chan et al.[31] In this model there are two sequential mechanisms for the relaxation. First, a rapid viscoelastic relaxation of the elastomer network, characterized by the viscoelastic timescale ($\tau_v$) and the dispersion factor ($\beta$). Then, a poroelastic relaxation due to the transport of oil away from the contact region within the elastomer described by its own characteristic timescale ($\tau_P$). In **Eq. (2)**, $F_0$ is the force at $t = 0$. $F_V$ and $F_P$ are the force after viscoelastic and poroelastic relaxation, respectively, obtained by fitting **Eq. (2)** to the PRI data.

$$F(t) = (F_0 - F_V)\,exp\left(-\frac{t}{\tau_V}\right)^\beta + (F_V - F_P)exp(-t/\tau_P) + F_P. \tag{2}$$

We also calculate the effective diffusivity, $D_{eff}$, of the silicone oil within the PDMS network during relaxation from:

$$\tau_P = \frac{R\delta}{D_{eff}}, \tag{3}$$

where $\sqrt{R\delta}$ is the radius of the contact region. Therefore, by increasing the radius of curvature R or indentation depth $\delta$, we can increase $\tau_P$. The fit for each relaxation curve to **Eq. (2)** is shown in **Fig. 2**, and the values of the fitting parameters as well as $D_{eff}$ are given in **Table 2**. For $\delta \sim 35\ \mu m$, $\tau_P$ is O(100-1000s).

For the case of dry extracted PDMS, the model in **Eq. (2)** reduces to a simple stretched exponential (**Eq. (4)**) that is useful for describing systems with a wide distribution of timescales. In literature both stretched exponentials and a Prony series have been used to describe the viscoelastic relaxation of crosslinked PDMS. [66, 67]

$$F(t) = (F_0 - F_V)\,exp\left(-\frac{t}{\tau_V}\right)^\beta + F_V \tag{4}$$

By fitting the data for dry PDMS to **Eq. (4)**, we get $\tau_V \sim$ O(1s) and $\beta \sim 0.4$. We can calculate the longest timescale over which most of the viscoelastic relaxation happens as a function of $\beta$ using the result by Johnston in 2006.[68] We see that the overall viscoelastic relaxation of dry PDMS is over by $\sim 100\ s$ in agreement with literature reports.[66, 69] We also observe that both swollen and dry PDMS have similar values of $\tau_V$. The quantity $\beta$ gives a measure of the spread around the characteristic timescales.[68] Values of $\beta < 1$ indicate an increase in the time necessary for the viscoelastic relaxation. Therefore, overall time of viscoelastic relaxation is larger for extracted and partially swollen PDMS than for fully swollen PDMS (**Fig. 2**).

We approximate the instantaneous relationship between the force during dwell $F$ and effective reduced Young's modulus ($E^* = \frac{E}{1-v^2}$; where $v$ is the Poisson ratio) of the swollen elastomer using Hertzian contact mechanics:

$$F = \frac{4}{3}E^* R^{0.5} \delta^{1.5}, \tag{5}$$

From value of $F_0, F_V$ and $F_P$, and substituting them into **Eq. (5)**, we obtain the initial effective modulus $E_0^*$, the effective modulus after viscoelastic relaxation $E_V^*$, and effective modulus after poroelastic relaxation $E_P^*$ (**Table S1**, ESI†). Relying on $E_V^*/E_0^*$ to characterize the extent of viscoelastic relaxation, we find that it is highest for dry PDMS and decreases with increasing oil fraction. Oil molecules occupy voids in the elastomer network and alter the extent to which the network chains can relax, limiting the capacity for viscoelastic relaxation.[70] For the fully swollen elastomer, $\frac{E_V^*}{E_0^*} \approx 1$ indicating little to no viscoelastic relaxation (**Table 2**). This limited viscoelastic relaxation is corroborated by oscillatory rheology as a decrease in $\tan \delta$ with increasing swelling (**Fig. S3**, ESI†). A similar loss of viscoelasticity has also been observed previously for swollen elastomers.[70, 71]

Extracted PDMS does not relax beyond the viscoelastic step, as expected (**Fig. 2**). For the two swollen PDMS $\tau_P \gg \tau_V$ at the given indentation depth, validating our use of **Eq. (2)**. While poroelastic behavior of silicone elastomer saturated with alkanes has been studied before,[33] silicones swollen in silicone oil have not been characterized via PRI and **Eq. (2)** has previously only been applied to saturated networks submerged in fluid. We find here that this model can describe the relaxation of both unsubmerged ($\phi = 0.4$ in air) and unsaturated networks ($\phi = 0.1$ in air). Data for fully swollen PDMS submerged in silicone oil is also shown as a control to get an estimate for what the effective diffusivity would be under submerged conditions. We find that the effective diffusivity $D_{eff}$ obtained is comparable for the three swollen cases. **Table 2**. $D_{eff}$ is a lumped parameter accounting for two fluid transport phenomena occurring simultaneously in the bulk of the elastomer- diffusion and convection. The dominant transport mechanism is determined by the relative sizes of the elastomer mesh and fluid molecule.[33] Here, the mesh size of the elastomer is ~2.6 nm.[60] An upper bound on the estimate for the size of the silicone oil molecule is $\sim 1\ nm$.[72] Since the mesh size is similar to the size of silicone oil molecule, the dominant contribution to $D_{eff}$ is likely from diffusive transport. Since the value of $D_{eff}$ are nearly the same for fully swollen PDMS in air and in an oil bath, we can assume that the mechanism for relaxation is the same regardless of the presence of an oil bath. We compare the values of $D_{eff}$ obtained from our PRI experiments with those from free swelling measurements done by Sotiri et al. using a linear poroelastic model (with $D \sim 1.2 \times 10^{-10}$ m²/s), which is comparable with the values in **Table 2**.[3] Our measurements also lie in the linear poroelastic regime due to the small deformation of the elastomer probe, thus validating our poroelastic characterization.

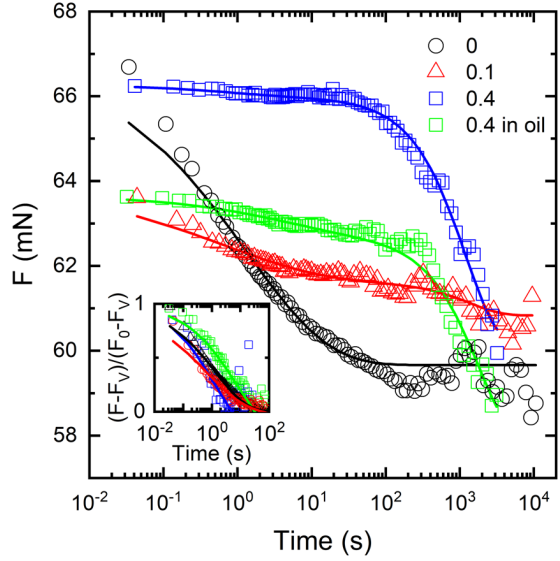

**Fig. 2. Representative force-time poroelastic relaxation indentation (PRI) curves for PDMS lens in contact with glass surface.** The different data sets correspond to different bulk oil fractions: $\phi = 0$ (black), $\phi = 0.1$ (red), $\phi = 0.4$ in air (blue) and $\phi = 0.4$ in silicone oil (green). All the curves are for the same indentation depth $\delta \approx 35\ \mu m$. Lines represent the numerical fit to Eq (2). Inset shows normalized force vs time curves for the first 100 s. The collapse of all the curves in the inset onto a single line at early times indicates similar viscoelastic timescales.

**Table 2.** Experimental and fitting parameters from PRI of dry and swollen PDMS.[1]

| $\phi$ | $\delta$ | $\tau_V$ | $\beta$ | $\tau_P$ | $E_V^*/E_0^*$ | $E_P^*/E_V^*$ | $D_{eff}$ |
|---|---|---|---|---|---|---|---|
| | mm | s | | s | | | x$10^{10}$ m$^2$/s |
| 0 | 33 ± 3 | 1.4 ± 0.5 | 0.4 ± 0.1 | -- | 0.83 ± 0.04 | -- | -- |
| 0.1 | 35 ± 3 | 1.4 ± 1.5 | 0.4 ± 0.1 | 285-3131 | 0.95 ± 0.01 | 0.97 ± 0.04 | 3.5 ± 2.7 |
| 0.4 | 36 ± 3 | 5.0 ± 4.2 | 0.7 ± 0.3 | 517-1478 | 0.99 ± 0.01 | 0.84 ± 0.16 | 2.2 ± 1.2 |
| (in oil) 0.4 | 38 ± 5 | 2.7 ± 1.3 | 0.6 ± 0.2 | 994-1829 | 0.99 ± 0.01 | 0.9 ± 0.01 | 2.0 ± 0.8 |

[1]values are averaged over three repeat experiments and errors reported are the standard deviations.

## B. Elastomer surface properties

Imaging of the PDMS surface shows the presence of oil droplets on the surface, even if the elastomer is not fully saturated with oil (**Fig. 3a**). The circular patches in the image are the droplets distributed evenly over the surface via autophobic dewetting of thin oil film,[73-75] and the remaining area is assumed to be dry polymer.[76] This dewetting behavior is due to an imbalance between the viscous and capillary forces acting on the oil film.[77] For fully swollen elastomers, we no longer see droplets, and expect the PDMS surface to be covered in a thin layer of oil. Prieto-Lopez *et. al.*[15] measured the equilibrium thickness of the oil surface to be greater than 100 nm. We anticipate that the presence of oil droplets at the surface of the swollen elastomer will impact its adhesion and wetting properties.

We characterized the surface energy of the elastomer at different oil fractions using contact angle measurements (**Fig. 3b**). The surface energy of the PDMS increases as the amount of oil dissolved within the matrix increases. We normalize the oil fraction $\phi$ by oil fraction at maximum swelling $\phi_{max} = 0.4$ in order to get the extent of saturation $\phi/\phi_{max}$. The dry extracted elastomer ($\phi/\phi_{max} = 0$) has a lower surface energy ($\gamma = 16.2 \pm 2$ mJ/m$^2$) compared to the surface tension of silicone oil (20.1 mJ/m$^2$). At $\phi = 0.2$, significantly before the saturation of the PDMS, the value of the surface energy of the PDMS surface saturates at $\gamma \sim 20.7 \pm 0.3$ mJ/m$^2$ (**Fig. 3b**). This value is close to the surface tension of silicone oil. The fact that the surface energy reaches the limit of the surface tension of the pure oil is indicative of a surface fully covered with oil. Therefore, for $\phi > 0.2$ the surface is fully saturated with oil while the bulk is only 50% saturated (i.e., $\phi/\phi_{max} = 0.5$). Excess of oil at the interface of swollen elastomers is known as syneresis, and occurs due to an imbalance between the mixing energy of the oil in the elastomer and the surface energy of the system.[4, 78] As the surface energy of the swollen elastomer increases with oil fraction we conclude that the oil at the interface contributes to the increasing surface energy of the elastomer.

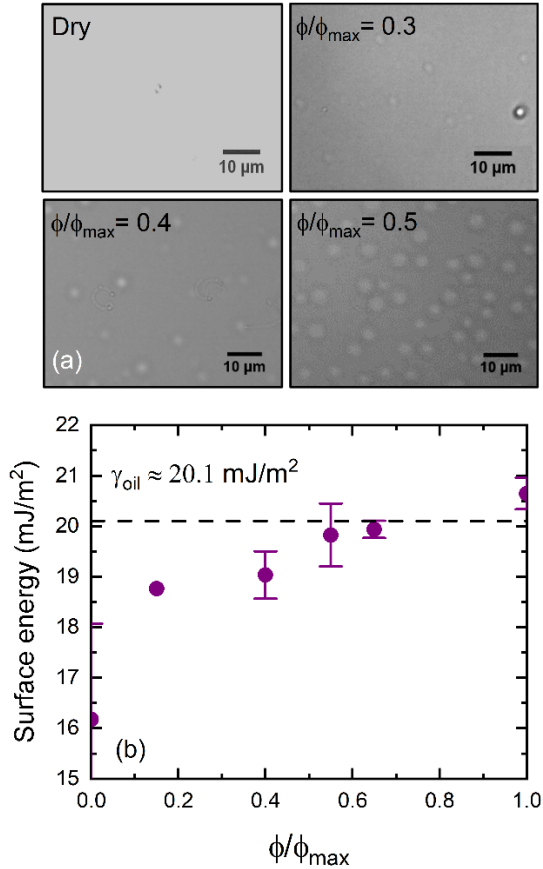

**Fig. 3. Surface characterization of swollen PDMS.** Flat PDMS sheets are swollen to different oil fractions from $\phi = 0$ (dry) to $\phi = \phi_{max} = 0.4$ (fully saturated). (a) Images of the sheets in air using reflectance microscopy at different normalized oil fraction, $\phi/\phi_{max}$. The circular spots are silicone oil droplets covering the surface, and the remainder of the surface is assumed to be dry (or oil-free due to dewetting). (b) Surface energy from the two-liquid method as a function of normalized oil fraction, $\phi/\phi_{max}$.

### C. Adhesion to swollen PDMS

We investigate how adhesion varies with oil fraction and contact time. We first look at adhesion data at contact times of 100 s for a wide range of oil fractions. We select a contact time of 100 s because at that time the viscoelastic relaxation of the PDMS is over and the impact of poroelasticity is still negligible, allowing us to isolate the effect of swelling. We then look at the effect of contact time on adhesion for three oil fractions across both viscoelastic and poroelastic timescales.

Role of oil fraction on adhesion (contact time 100 s)

The oil fraction has a strong influence on adhesion (**Fig. 4a**). The adhesive strength (peak force in **Fig. 4a**) during pull-off decreases rapidly with an increase in oil fraction. This decrease is consistent with the recent reports of Ibáñez-Ibáñez et al.[24] where they see a similar sharp decrease in ice adhesion in tensile strength measurements. For comparison, our nominal stress ($\sigma = F_P/A$) reduces by a factor of four between dry and 23% saturated PDMS (i.e., $\phi/\phi_{max}$ ~0.23). Similarly, ice adhesion strength also decreases by a factor of four between dry and 20% saturated PDMS. While in the ice adhesion measurements a column of water is frozen over the swollen elastomer, here we simply bring the two surfaces in contact at room temperature. It is interesting to note that the protocol for contact formation (ice or glass) does not seem to affect the relative effect of oil on adhesion meaningfully. This similarity could indicate that the amount of oil at the interface is a strong determinant of the adhesion of swollen PDMS with other surfaces independent of the surface properties of the substrate (e.g., glass or ice).

For fully-swollen PDMS we observe a tail in the force-time curve and the presence of a large oil droplet on the glass surface after detachment (**Fig. 4b**). Side view imaging during debonding also shows the presence (and break up) of a capillary bridge between the surfaces (**Fig. S7**, ESI†)**.** We therefore hypothesize that the detachment force is due to both capillary and solid-solid contact. We developed a simple model to describe the debonding curve where the total force at any time during debonding, $F_{tot}$ , is given by a sum of the capillary ($F_{cap}$) and an adhesive ($F_{JKR}$) forces, see **Fig. 4c** and **Fig. S8**, ESI† for a description. We include solid-solid adhesion via JKR mechanics to obtain a theoretical upper bound on the total force. Comparison with the model shows that the force-time curve has a contribution of both solid-solid contact and capillarity, but contribution of solid-solid contact is negligible compared to the unsaturated case. The contribution of solid-solid contact is unexpected considering the fact that contact angle measurements indicated that the PDMS surface was initially covered with an oil film (**Fig. 3b**). In contrast, while contact angle measurements showed a significant portion of the surface covered with oil even before saturation (**Fig. 3**), we do not observe a tail in the force curves or visualize any capillary bridges during detachment for unsaturated PDMS. It is possible that small capillary bridges might be present but their contribution to the adhesive strength is negligible. For adhesion of fully-swollen PDMS, wetting of the glass surface by oil determines the low adhesion value seen in our experiments.

For all swelling fraction, we convert the adhesive strength $F_P$ into a critical strain energy release rate ($G_C$), a quantity that only depends on the interfacial properties of the materials in contact and the rate of crack propagation (and exclude effects due to geometry or compliance). The critical strain energy release rate is obtained using the JKR relationship for sphere plane contact given by **Eq. (6)**[79]

$$F_P = \frac{3}{2}\pi R G_C \qquad (6)$$

where $F_P$ is the peak force during detachment. Since we debond at the same velocity for all experiments ($v_{debond} = 50 \mu m/s$), we assume that rate dependence of $G_C$ on the crack propagation will be similar across different oil fraction (see ESI†). Therefore, $G_C$ allows for comparison across oil fractions that are due to changes in the interfacial properties of the swollen elastomer in contact with glass substrate.

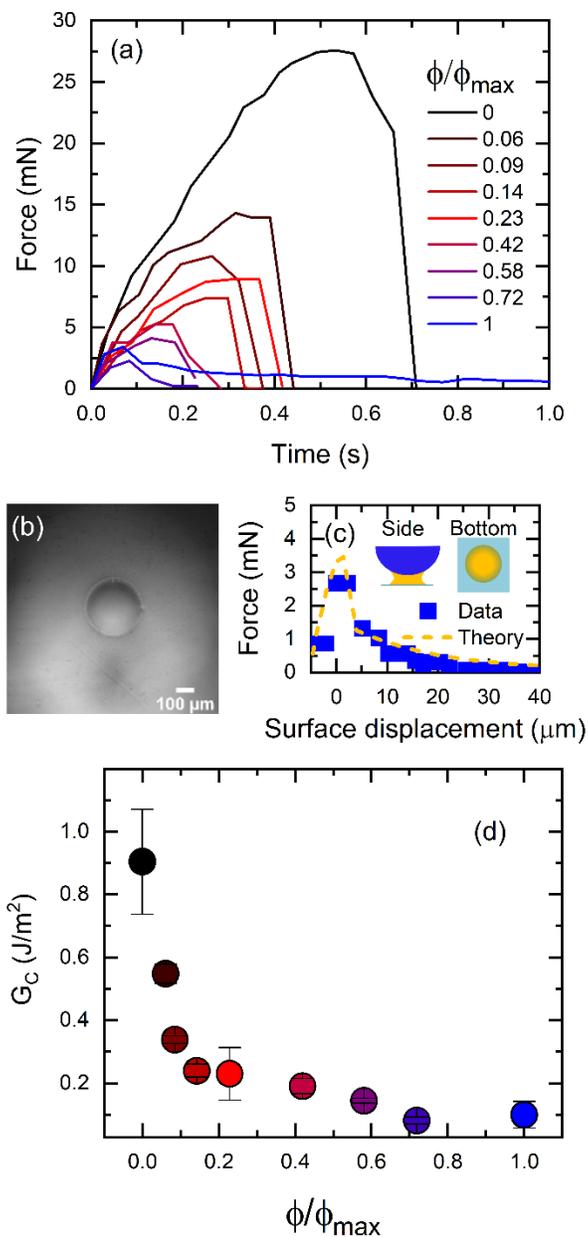

**Fig. 4. Short dwell adhesion of glass-PDMS as a function of the PDMS oil fraction.** (a) Representative force vs time curves for debonding of swollen PDMS lenses from flat glass with different normalized oil fraction, $\phi/\phi_{max}$ (b) Image of a silicone oil droplet on glass taken after debonding from a fully-saturated PDMS lens. (c) Force vs time data for debonding of fully saturated PDMS ($\phi_{max} = 0.4$) from glass. Blue points are data and the dashed yellow line is a theoretical calculation for an elastic lens debonding in the presence of a capillary bridge[60]. (d) Critical strain energy release rate $G_C$ as a functions of normalized bulk oil fraction, $\phi/\phi_{max}$. $G_C$ decreases by a factor of 10 as the oil fraction in PDMS increases.

We attempt to compare adhesion and contact angle data to contextualize the effect of oil fraction on adhesion. To do so we calculate an effective surface fraction of oil for both the contact angle and adhesion measurements as $\phi_s$. This effective surface fraction depends on the oil fraction, and is obtained from the energy of the pure components assuming that the surface energy of the oil-covered PDMS follows Cassie's relation:[80, 81]

$$\gamma_{calc} = \gamma_{oil}\,\phi_s + \gamma_{dry}\,(1-\phi_s), \tag{7}$$

where $\gamma_{calc}$ is the surface energy of the swollen surface in air; $\gamma_{oil}$ and $\gamma_{dry}$ are the surface tension of the fluid and surface energy of the dry PDMS in air obtained using the two-liquid method. $\phi_s$ is the area fraction of the surface covered by the silicone oil and varies with the oil fraction. Given that the maximum surface fraction of oil is $\phi_{s,max}=1$ (for a surface covered with a layer of oil), we can treat surface fraction $\phi_s$ as the extent of saturation at the surface, $\frac{\phi_s}{\phi_{s,max}}$. From **Fig. 5**, it can be seen that the surface is always more saturated than the bulk ($\phi_s > \frac{\phi}{\phi_{max}}$).

We follow a similar mixing rule as for the contact angle to estimate an effective area fraction of oil within the contact region during adhesion measurements. We assume that $G_C$ is a sum of the contribution from PDMS-glass interactions and oil-glass interactions as given by **Eq. (8)**

$$G_{C,meas} = G_{C,oil-glass}\,\phi_s + G_{C,dry-glass}(1-\phi_s). \tag{8}$$

Data points in **Fig. 5** are a rescaling of the data presented in **Figs. 3-4**. Cassie's relation has been applied previously to estimate the adhesion on patterned surfaces as a function of area fraction.[82] Since the adhesion between fully swollen PDMS and glass $G_{C,oil-glass}$ is negligible compared to adhesion between dry-PDMS and glass $G_{C,dry-glass}$, we can assume that the contact formed will be similar to Cassie-Baxter state. The $\phi_s$ calculated from adhesion data is always higher than the one obtained for a free surface using contact angle data. As a result, **Fig. 5a** can be interpreted as the difference in the state (in terms of oil area fraction) of the elastomer surface before and during contact, as illustrated in **Fig. 5b**. The increase in $\phi_s$ during contact could come from spreading of the oil droplets within the contact region resulting in higher coverage of oil. Interestingly, while oil volume fraction greater than $\frac{\phi}{\phi_{max}} \sim 0.2$ show small in change effective oil surface fraction (it is almost fully covered) the energy release rate continues to decrease. Moreover, capillary bridges are only observed (visually or through a tail in the force-time curves) for fully saturated PDMS lenses. These observations indicate that to achieve low adhesion, the elastomer must be swollen above a minimum threshold oil fraction, likely fully saturated.

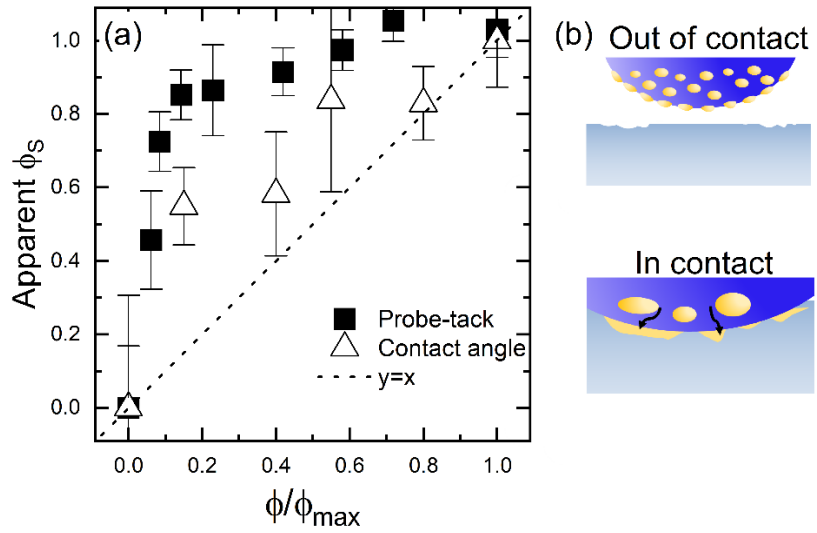

**Fig. 5. Surface distribution of oil on free surface and surface in contact.** (a) Apparent surface fraction of oil $\phi_s$ vs normalized bulk oil fraction $\phi/\phi_{max}$ calculated by applying Cassie's law for heterogenous surfaces, Eqns. (7-8). Triangles are calculated using surface energy obtained from contact angle measurements. Squares are calculated using strain energy release rates from probe-tack measurements. The surface has an excess of oil (all points are above $y = x$ line) in both for both adhesion and wetting. The apparent surface fraction of oil "felt" by the glass surface is greater than that obtained from contact angle measurements. The apparent surface fraction in contact saturates around $\frac{\phi}{\phi_{max}}$ ~0.14. (b) Schematic of oil distribution before /after the lens comes in contact with the glass slide. Yellow coloured regions represent oil at the interface. The small grooves on the flat glass surface represent local roughness of glass

Contact time dependent adhesion

We then compare the effect of contact time on the adhesion of dry, unsaturated PDMS ($\phi = 0.1$), and fully-swollen PDMS. The oil fraction of $\phi = 0.1$ was also selected as it is just below the oil fraction above which we report an oil film covering the entirety of the contact region. The contact times selected span the whole relaxation time scale shown in **Fig. 2**. Our objective is to determine the relative importance of three different mechanisms for time-dependent adhesion: 1) viscous dissipation within the probe prior to debonding,[83, 84] 2) enhancement of solid-solid contact (contact aging),[85, 86] and 3) suction caused by fluid transport away from the contact region (poroelasticity).[45, 46]

For all contact times, the debonding force is the highest for dry PDMS probe and lowest for the fully swollen probe. For all contact times we see a tail in the force curve for the fully swollen probe due to capillary bridges formed during debonding (**Fig. 6c**). The representative force-time curves shown in **Fig.**

6a-c show that the peak force increases with contact time for $\phi = 0$ and $\phi = 0.1$, but much less so (if any) for the fully-saturated PDMS.

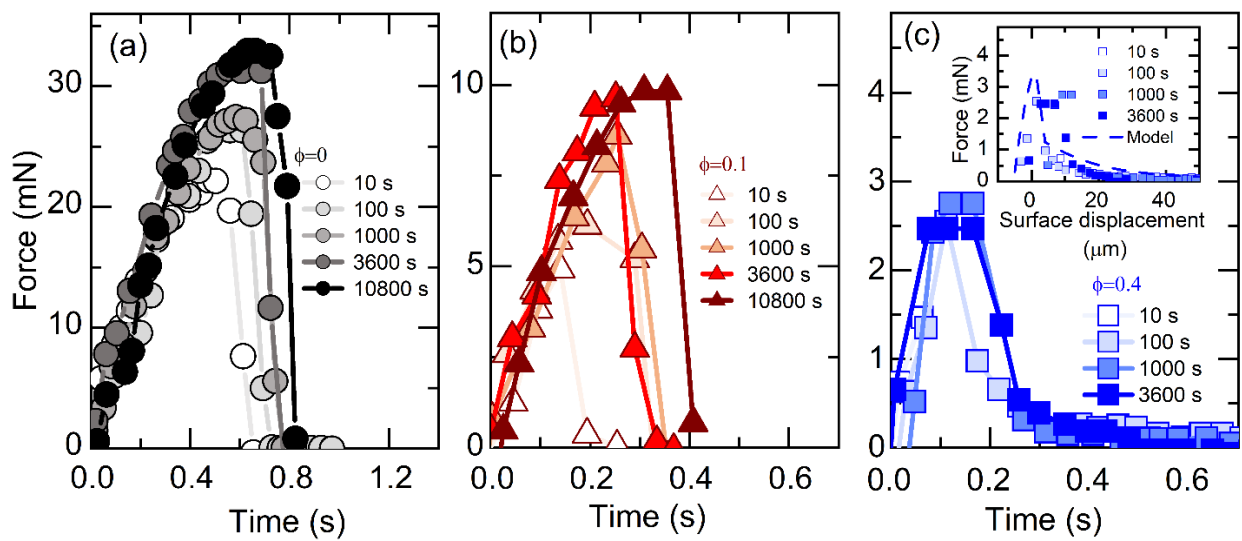

**Fig. 6. Representative force vs time curves during debonding of PDMS lens at different contact times.** (a) $\phi = 0$ or dry PDMS (black circles), (b) $\phi = 0.1$ or partially swollen PDMS (red triangles) and (c) $\phi = 0.4$ or fully saturated PDMS (blue squares). Data at increasing contact times is shown using darker shades. The peak force increases with contact time for $\phi = 0$ and $\phi = 0.1$. Inset in (c): force vs surface displacement. Dashed lines are calculated from the capillary/JKR adhesion model (calculations in ESI†). Fully swollen probes do not show a strong (if any) dependence of adhesion with contact time. Lines joining the points are to guide the eye.

Elastomeric materials exhibit viscoelastic properties which affect their adhesive behavior.[83, 84, 87-89] **Fig. 7a** shows the effective modulus $E^*$ (obtained using **Eq. (5)**) scaled by the effective viscoelastic modulus $E_V^*$ as a function of contact time. We see that $E^*/E_V^*$ decreases significantly until viscoelastic relaxation is complete (~100 s). Therefore, adhesion prior to 100 s will be sensitive to viscoelastic properties of the material. At timescales longer than those relevant to viscoelastic relaxation, a slow enhancement of solid-solid contact – known as contact aging – also contributes to increasing adhesion of the elastomer. While the exact mechanism behind contact aging is not well-established, it has been suggested that rearrangement of chains at the interface allows for better bonds to be formed.[41, 42, 90, 91] For PDMS in contact with a hydrophilic surface, the rearrangement is facilitated by the presence of active -OH groups on the surface of glass due to strong dipole-dipole interactions which in turn leads to an increase in adhesion.[91] Contact aging increases adhesion with a power law dependence on contact time $\sim t^n$ where n is a power law exponent that depends on the elastomer.[85, 86, 88, 90-92] Therefore, solid-solid adhesion for dry (extracted) elastomer can increase with time due to both increased viscous dissipation in the bulk of the elastomer and contact aging at the interface. Typically, the timescale for the viscoelastic relaxation is much shorter than for contact aging. For swollen hydrogels, it has also been seen that the increase in adhesion with contact

Jha et al. Page 17 of 37

time occurs beyond the poroelastic time $\tau_P$ measured via indentation relaxation.[45] As a result, it is speculated that a "suction" pressure develops across the interface of the hydrogel due to fluid drainage in the bulk near the interface leading to an increase in adhesion.[45, 46, 48, 50] Reale et al.[45] obtained a time dependent surface energy of the indented hydrogel under water and observed that the increase happened over the poroelastic timescale. McGhee et al. report that the increase in adhesion comes from an interfacial stress gradient that is created due to a difference in the osmotic pressure in the hydrogel and the hydrated second surface.[46]

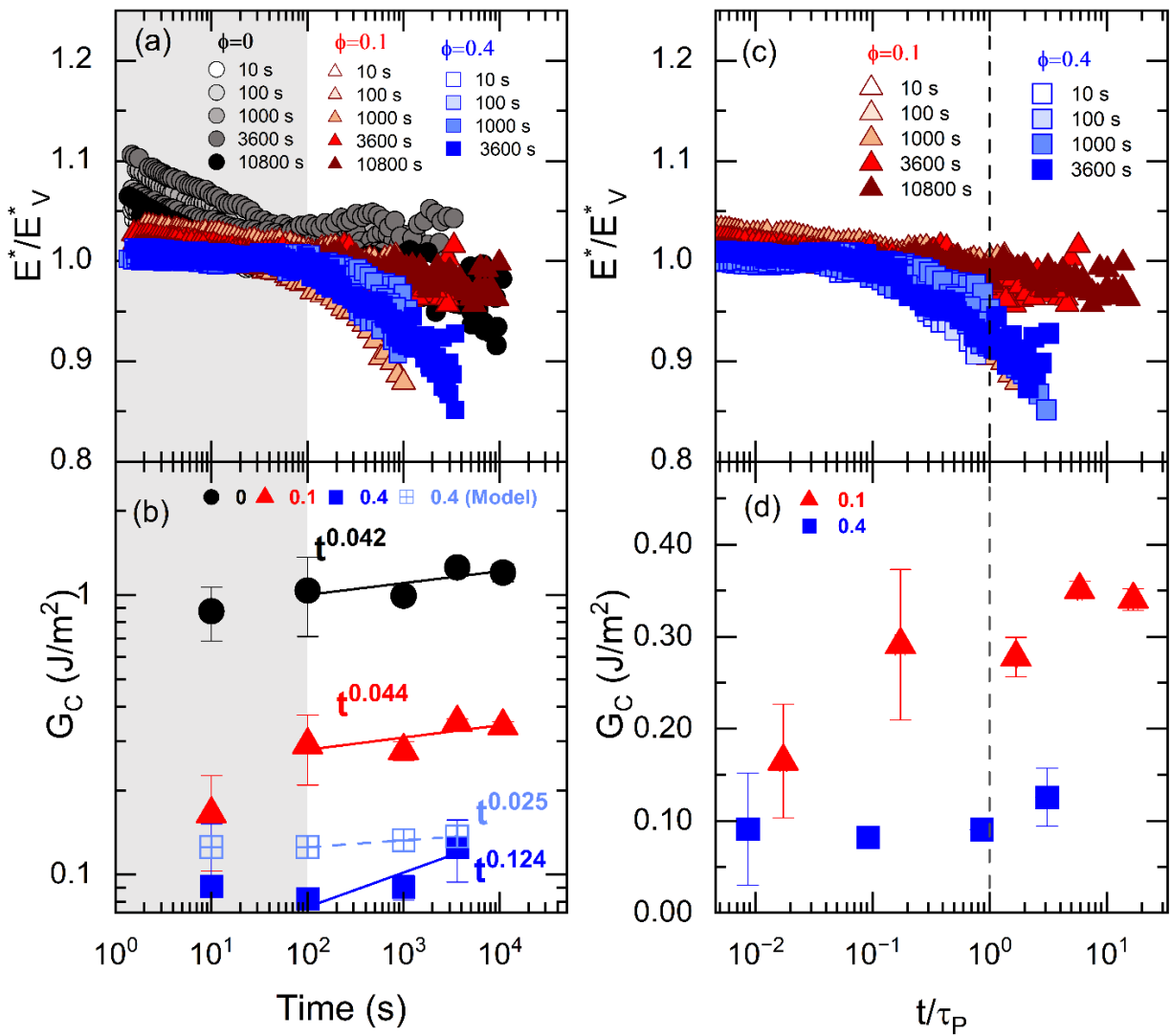

**Fig. 7. Effect of contact time on relaxation and adhesion of PDMS for different oil fraction.** (a, b) Normalized relaxation modulus ($E^*/E_V^*$ from Eq. (5)) and critical strain energy release rate ($G_C$) vs. contact time. The shaded area shows the regime for viscoelastic relaxation (c, d) Normalized relaxation modulus and critical strain energy release

rate vs. contact time normalized with poroelastic timescale, $t/\tau_P$, for each experiment where $\tau_P$ is obtained by fitting relaxation data to Eq. (2). The dashed line represents the poroelastic timescale $t/\tau_P = 1$ Each set of data represent the oil fraction with $\phi = 0$ or dry (black circles), $\phi = 0.1$ or partially swollen (red triangles) and $\phi = 0.4$ or fully saturated (blue squares). (b) The black, red, and blue solid lines represent power law dependence in time. Blue dashed line indicates magnitude of $G_C$ for capillary adhesion limit obtained from Eq (5) using peak force calculated from capillary/JKR adhesion model[60]. The data is plotted on a log scale on the x and y axes to highlight power-law dependence and similar slope for $\phi = 0$ and $\phi = 0.1$ (d) Data is plotted on a linear scale in $G_C$ to compare the magnitude of $G_C$ for the partially ($\phi = 0.1$) and fully swollen ($\phi = 0.4$) cases. Data points for $\phi = 0$ are not plotted because there is no characteristic poroelastic timescale for extracted PDMS.

For dry PDMS in contact with glass, there is an initial increase in adhesion (**Fig. 7b**) which we attribute to viscous dissipation in the bulk. We see in **Fig. 7a** that viscoelastic relaxation lasts until $\sim 100\ s$. Therefore, initial increase in $G_C$ from $\tau_D = 10\ s$ to $\tau_D = 100\ s$ is due to changing viscoelastic properties of the material. Beyond 100 s, the contribution of bulk viscoelastic dissipation to adhesion are insignificant. At longer times, we attribute further increase in adhesion to contact aging and obtain a power law dependence for $G_C$ on contact time of $\sim t^{0.04}$, which is lower than $\sim t^{0.1}$ previously reported for work of debonding.[40, 86]

For elastomer fully saturated with fluid, adhesion measurements at all contact times lie in a regime where force due to liquid capillary bridge meniscus is comparable to measured debonding force (**Fig. 6c**). There is poor solid-solid contact: the peak force during detachment is an order of magnitude lower for $\phi = 0.4$ than the peak force for dry case. Because of low adhesion we expect that viscoelastic dissipation and contact aging will not contribute significantly to increase in adhesion. We observe a slight increase in adhesion across the poroelastic timescale shown in the scaling for **Fig. 7d**, however the increase(0.04 J/m$^2$) is low enough to lie within limits of experimental resolution(0.05 J/m$^2$). Comparison with our model for capillary and JKR adhesion shows that both the liquid meniscus and solid-solid contact contribute to the detachment force **(Fig. S8, S9**, ESI†). We calculate a strain energy release rate just for the capillary contribution, $G_{C,CAP}$ using **Eq. (6)** and the peak force from the model. Although $G_C$ is used to describe crack propagation in solids, we extend this definition to breakup of liquid capillary meniscus just for comparison in terms of relative magnitude. Since the magnitude of adhesion $G_C$ is below the limit of capillary bridge break-up, $G_{C,CAP}$ (**Fig. 7c, d**) and that the increase in adhesion is negligible, we conclude that there is little to no contribution to poroelasticity induced suction to increasing adhesion for fully swollen PDMS. The absence of a poroelastic contribution could be due to insufficient buildup of suction pressure that would drive the transport of the fluid and increase adhesion. Asperities on the glass surface could lead to a poor "seal" which would work against pressure buildup near the interface. As a result, fluid trapped at the interface[93] would not be taken up by the elastomer leading to consistently poor adhesion. Another reason for poor pressure buildup could be due to the poor permeability of silicone oil through the medium under compression perhaps due to the comparable size of the fluid molecule and the elastomer mesh size. The total stress in a poroelastic material is a sum of the elastic and fluid contribution.[94] It is also possible that for our material, the elastic contribution to stress far outweighs the fluid contribution leading to negligible effect of poroelasticity.

For the partially swollen PDMS ($\phi = 0.1$) we see that the pull-off forces increase with contact time. Even for the longest time in contact the pull-off force remains significantly lower than the one measured with the dry PDMS probe (**Fig. 6**). There is negligible change in the time dependence of $G_C$ across the poroelastic timescale ($G_C = 0.29 \pm 0.08 - 0.27 \pm 0.02$ J/m²). At longer times, we also do not recover the dry value of adhesion, likely due to fluid wetting and entrapment near the interface. We also see the same power law dependence ($t^{0.044}$) as the dry probe-glass adhesion for the partially swollen probe-glass adhesion. A same power law dependence supports the hypothesis that the increase in adhesion is due to contact aging. We have two features of the adhesion data to suggest that solid-solid contact formation and aging determine the adhesion of partially swollen elastomers: (1) similar power law increase in adhesion ($G_C$) with time for both dry and swollen lenses and (2) no distinction in the increase in adhesion across poroelastic timescale for the partially swollen elastomers. For both the dry and fully swollen PDMS, we see that further relaxation beyond viscoelastic relaxation (see in **Fig. 7c**) does not contribute to increasing adhesion (**Fig. 7d**).

**CONCLUSIONS**

The main objective of this work was to understand the effect of swelling and contact time on fluid-infused elastomers. We did so by systematically varying the amount of oil in the bulk of the PDMS elastomer and measuring its surface and rheological properties. Contact angle measurements showed an excess of oil at the elastomer droplet interface. The oil at the surface contributes to the increasing surface energy of the elastomer – reaching the value of pure oil at complete saturation. Simple brightfield imaging of the surface showed the presence of oil droplets at the interface that can be explained via autophobic dewetting of the thin film of oil. We saw a loss of viscous behavior with increasing higher oil fractions in both rheology and PRI measurements, which could be due to filling up of voids in the elastomer at higher oil fractions.

Adhesion decreases sharply with increasing oil content in the system – reaching the limit of capillary adhesion at full saturation. Using a Cassie-Baxter type linear relationship, we obtained an effective surface fraction from the surface energy and strain energy release rate of the swollen elastomer. The effective surface fraction of oil reaches its maximum value ($\phi_s = 1$) earlier during adhesion than during contact angle measurements. This can be due to the spreading of the oil droplets and wetting of the glass. A higher effective surface fraction during contact indicates that the surface gets more "slippery" in contact with glass as compared to liquid droplets. It remains to be seen how the surface would behave in the presence of a softer, polymeric contacting surface as compared to the rigid glass surface studied here.

We also looked at the effect of contact time on the adhesion of dry and swollen elastomers with the glass surface. Both dry and swollen elastomers undergo visco-poroelastic relaxation, which can be observed during PRI tests. During debonding, adhesion of dry ($\phi = 0$) and partially swollen ($\phi = 0.1$) lenses follows a similar power law dependence in time, with an exponent of $n = 0.04$. This value is slightly lower than the value of $n = 0.1$ seen for dry PDMS elastomers. We hypothesize that the similar dependence of adhesion with time originates from similar adhesive mechanism – i.e., contact aging. We also consider poroelasticity as a possible mechanism for increase in adhesion via fluid diffusion away from the interface, but do not see a significant change in the time dependence across the characteristic poroelastic timescale for both the partially swollen and fully swollen elastomer. We believe that the "suction" effect, if present, is not significant enough to deplete the interface of oil and increase adhesion. It is possible that more intimate contact formation is required - which is possible with softer materials to develop and maintain a pressure gradient. For the fully swollen elastomer, even though the diffusion coefficient is similar from PRI in air and submerged in oil, effect of change in boundary condition (flooded vs. unflooded) on adhesion is yet to be investigated and shall be the subject of a future study. Our results indicate that the time dependent increase in adhesion for PDMS is determined by the initial solid-solid contact which is determined by the amount of oil present at the surface of the elastomer. The elastomer needs to be swollen above a threshold oil fraction in order to maintain low adhesion performance over time.

## ACKNOWLEDGEMENTS

This work was partially supported by the National Science Foundation through NSF-CMMI 1728082 and ACS-PRF 58606-ND5.

## REFERENCES

1. J. Li, E. Ueda, D. Paulssen and P. A. Levkin, Slippery Lubricant-Infused Surfaces: Properties and Emerging Applications, *Adv. Funct. Mater.*, 2019, **29**, 1802317.
2. Z. Hao and W. Li, A Review of Smart Lubricant-Infused Surfaces for Droplet Manipulation, *Nanomaterials*, 2021, **11**, 801.
3. I. Sotiri, A. Tajik, Y. Lai, C. T. Zhang, Y. Kovalenko, C. R. Nemr, H. Ledoux, J. Alvarenga, E. Johnson, H. S. Patanwala, J. V. I. Timonen, Y. H. Hu, J. Aizenberg and C. Howell, Tunability of liquid-infused silicone materials for biointerfaces, *Biointerphases*, 2018, **13**, 06D401.
4. N. Lavielle, D. Asker and B. D. Hatton, Lubrication dynamics of swollen silicones to limit long term fouling and microbial biofilms, *Soft Matter*, 2021, **17**, 936-946.
5. K. Golovin and A. Tuteja, A predictive framework for the design and fabrication of icephobic polymers, *Sci. Adv.*, 2017, **3**, e1701617.

6.  J. Kim, M. K. Chaudhury, M. J. Owen and T. Orbeck, The Mechanisms of Hydrophobic Recovery of Polydimethylsiloxane Elastomers Exposed to Partial Electrical Discharges, *J. Colloid Interface Sci.*, 2001, **244**, 200-207.
7.  Y. H. Yeong, C. Y. Wang, K. J. Wynne and M. C. Gupta, Oil-Infused Superhydrophobic Silicone Material for Low Ice Adhesion with Long-Term Infusion Stability, *ACS Appl. Mater. Interfaces*, 2016, **8**, 32050-32059.
8.  J. H. Kim, M. J. Kim, B. Lee, J. M. Chun, V. Patil and Y.-S. Kim, Durable ice-lubricating surfaces based on polydimethylsiloxane embedded silicone oil infused silica aerogel, *Appl. Surf. Sci.*, 2020, **512**, 145728.
9.  J. Zhang, B. Liu, Y. Tian, F. Wang, Q. Chen, F. Zhang, H. Qian and L. Ma, Facile One-Step Method to Fabricate a Slippery Lubricant-Infused Surface (LIS) with Self-Replenishment Properties for Anti-Icing Applications, *Coatings*, 2020, **10**, 119.
10. L. Zhao, L. He, J. Liang, Y. Chen, M. Jia and J. Huang, Facile preparation of a slippery oil-infused polymer surface for robust icephobicity, *Prog. Org. Coat.*, 2020, **148**, 105849.
11. C. Wang, M. C. Gupta, Y. H. Yeong and K. J. Wynne, Factors affecting the adhesion of ice to polymer substrates, *J. Appl. Polym. Sci.*, 2018, **135**, 45734.
12. A. T. Mulroney and M. C. Gupta, Ice adhesion properties of periodic surface microtextured optically transparent superhydrophobic polydimethylsiloxane, *Cold Reg. Sci. Technol.*, 2022, **198**, 103540.
13. G. Liu, Y. Yuan, R. Liao, L. Wang and X. Gao, Fabrication of a Porous Slippery Icephobic Surface and Effect of Lubricant Viscosity on Anti-Icing Properties and Durability, *Coatings*, 2020, **10**, 896.
14. S. Amini, S. Kolle, L. Petrone, O. Ahanotu, S. Sunny, C. N. Sutanto, S. Hoon, L. Cohen, J. C. Weaver, J. Aizenberg, N. Vogel and A. Miserez, Preventing mussel adhesion using lubricant-infused materials, *Science*, 2017, **357**, 668-673.
15. L. O. Prieto-Lopez, P. Herbeck-Engel, L. Yang, Q. Wu, J. T. Li and J. X. Cui, When Ultimate Adhesive Mechanism Meets Ultimate Anti-Fouling Surfaces-Polydopamine Versus SLIPS: Which One Prevails?, *Adv. Mater. Interfaces*, 2020, **7**, 2000876.
16. P. Zhang, C. Zhao, T. Zhao, M. Liu and L. Jiang, Recent Advances in Bioinspired Gel Surfaces with Superwettability and Special Adhesion, *Adv. Sci.*, 2019, **6**, 1900996.
17. J. D. Tang, J. Y. Li, J. J. Vlassak and Z. G. Suo, Adhesion between highly stretchable materials, *Soft Matter*, 2016, **12**, 1093-1099.
18. S. Boyadzhieva, K. Sorg, M. Danner, S. C. L. Fischer, R. Hensel, B. Schick, G. Wenzel, E. Arzt and K. Kruttwig, A Self-Adhesive Elastomeric Wound Scaffold for Sensitive Adhesion to Tissue, *Polymers*, 2019, **11**, 942.
19. S. C. L. Fischer, S. Boyadzhieva, R. Hensel, K. Kruttwig and E. Arzt, Adhesion and relaxation of a soft elastomer on surfaces with skin like roughness, *J. Mech. Behav. Biomed. Mater.*, 2018, **80**, 303-310.
20. W. Zhang, B. H. Wu, S. T. Sun and P. Y. Wu, Skin-like mechanoresponsive self-healing ionic elastomer from supramolecular zwitterionic network, *Nat. Commun.*, 2021, **12**, 4082.
21. M. Nachman and S. E. Franklin, Artificial Skin Model simulating dry and moist in vivo human skin friction and deformation behaviour, *Tribol. Int.*, 2016, **97**, 431-439.
22. J. Chen, H. Yang, J. Li, J. Chen, Y. Zhang and X. Zeng, The development of an artificial skin model and its frictional interaction with wound dressings, *J. Mech. Behav. Biomed. Mater.*, 2019, **94**, 308-316.

23. M. Saito, T. Nishimura, K. Sakiyama and S. Inagaki, Self-healing of optical functions by molecular metabolism in a swollen elastomer, *AIP Adv.*, 2012, **2**, 042118.
24. P. F. Ibáñez-Ibáñez, F. J. Montes Ruiz-Cabello, M. A. Cabrerizo-Vílchez and M. A. Rodríguez-Valverde, Ice adhesion of PDMS surfaces with balanced elastic and water-repellent properties, *J. Colloid Interface Sci.*, 2022, **608**, 792-799.
25. W. S. Y. Wong, L. Hauer, A. Naga, A. Kaltbeitzel, P. Baumli, R. Berger, M. D'Acunzi, D. Vollmer and H.-J. Butt, Adaptive Wetting of Polydimethylsiloxane, *Langmuir*, 2020, **36**, 7236-7245.
26. A. Hourlier-Fargette, A. Antkowiak, A. Chateauminois and S. Neukirch, Role of uncrosslinked chains in droplets dynamics on silicone elastomers, *Soft Matter*, 2017, **13**, 3484-3491.
27. Z. Cai, A. Skabeev, S. Morozova and J. T. Pham, Fluid separation and network deformation in wetting of soft and swollen surfaces, *Commun. Mater.*, 2021, **2**, 21.
28. Z. Cai and J. T. Pham, How Swelling, Cross-Linking, and Aging Affect Drop Pinning on Lubricant-Infused, Low Modulus Elastomers, *ACS Appl. Polym. Mater.*, 2022, **4**, 3013-3022.
29. J. D. Berman, M. Randeria, R. W. Style, Q. Xu, J. R. Nichols, A. J. Duncan, M. Loewenberg, E. R. Dufresne and K. E. Jensen, Singular dynamics in the failure of soft adhesive contacts, *Soft Matter*, 2019, **15**, 1327-1334.
30. K. E. Jensen, R. Sarfati, R. W. Style, R. Boltyanskiy, A. Chakrabarti, M. K. Chaudhury and E. R. Dufresne, Wetting and phase separation in soft adhesion, *Proc. Natl. Acad. Sci.*, 2015, **112**, 14490-14494.
31. E. P. Chan, B. Deeyaa, P. M. Johnson and C. M. Stafford, Poroelastic relaxation of polymer-loaded hydrogels, *Soft Matter*, 2012, **8**, 8234-8240.
32. Y. Lai, D. J. He and Y. H. Hu, Indentation adhesion of hydrogels over a wide range of length and time scales, *Extreme Mech. Lett.*, 2019, **31**, 100540.
33. Y. H. Hu, X. Chen, G. M. Whitesides, J. J. Vlassak and Z. G. Suo, Indentation of polydimethylsiloxane submerged in organic solvents, *J. Mater. Res.*, 2011, **26**, 785-795.
34. M. L. Oyen, Mechanical characterisation of hydrogel materials, *Int. Mater. Rev.*, 2014, **59**, 44-59.
35. Q. M. Wang, A. C. Mohan, M. L. Oyen and X. H. Zhao, Separating viscoelasticity and poroelasticity of gels with different length and time scales, *Acta Mech. Sin.*, 2014, **30**, 20-27.
36. Q. Xu, L. A. Wilen, K. E. Jensen, R. W. Style and E. R. Dufresne, Viscoelastic and Poroelastic Relaxations of Soft Solid Surfaces, *Phys. Rev. Lett.*, 2020, **125**, 238002.
37. Y. H. Hu, X. H. Zhao, J. J. Vlassak and Z. G. Suo, Using indentation to characterize the poroelasticity of gels, *Appl. Phys. Lett.*, 2010, **96**, 121904.
38. D. Okumura, H. Kawabata and S. A. Chester, A general expression for linearized properties of swollen elastomers undergoing large deformations, *J. Mech. Phys. Solids*, 2020, **135**, 103805.
39. D. Okumura, A. Kondo and N. Ohno, Using two scaling exponents to describe the mechanical properties of swollen elastomers, *J. Mech. Phys. Solids*, 2016, **90**, 61-76.
40. J. Thiemecke and R. Hensel, Contact Aging Enhances Adhesion of Micropatterned Silicone Adhesives to Glass Substrates, *Adv. Funct. Mater.*, 2020, **30**, 2005826.

41. A. N. Gent and R. H. Tobias, Effect of interfacial bonding on the strength of adhesion of elastomers. III. Interlinking by molecular entanglements, *J. Polym. Sci., Polym. Phys. Ed.*, 1984, **22**, 1483-1490.
42. R. Buzio, A. Bosca, S. Krol, D. Marchetto, S. Valeri and U. Valbusa, Deformation and Adhesion of Elastomer Poly(dimethylsiloxane) Colloidal AFM Probes, *Langmuir*, 2007, **23**, 9293-9302.
43. J. D. Berry, M. Biviano and R. R. Dagastine, Poroelastic properties of hydrogel microparticles, *Soft Matter*, 2020, **16**, 5314-5324.
44. B. G. Bush, J. M. Shapiro, F. W. DelRio, R. F. Cook and M. L. Oyen, Mechanical measurements of heterogeneity and length scale effects in PEG-based hydrogels, *Soft Matter*, 2015, **11**, 7191-7200.
45. E. R. Reale and A. C. Dunn, Poroelasticity-driven lubrication in hydrogel interfaces, *Soft Matter*, 2017, **13**, 428-435.
46. E. O. McGhee, S. M. Hart, J. M. Uruena and W. G. Sawyer, Hydration Control of Gel-Adhesion and Muco-Adhesion, *Langmuir*, 2019, **35**, 15769-15775.
47. Y. Qi, K. N. Calahan, M. E. Rentschler and R. Long, Friction between a plane strain circular indenter and a thick poroelastic substrate, *Mech. Mater.*, 2020, **142**, 103303.
48. L. Ciapa, J. Delavoipiere, Y. Tran, E. Verneuil and A. Chateauminois, Transient sliding of thin hydrogel films: the role of poroelasticity, *Soft Matter*, 2020, **16**, 6539-6548.
49. T. Shoaib and R. M. Espinosa-Marzal, Insight into the Viscous and Adhesive Contributions to Hydrogel Friction, *Tribol. Lett.*, 2018, **66**, 96.
50. G. Han and M. Eriten, Effect of relaxation-dependent adhesion on pre-sliding response of cartilage, *R. Soc. Open Sci.*, 2018, **5**, 172051.
51. J. Delavoipiere, Y. Tran, E. Verneuil, B. Heurtefeu, C. Y. Hui and A. Chateauminois, Friction of Poroelastic Contacts with Thin Hydrogel Films, *Langmuir*, 2018, **34**, 9617-9626.
52. G. K. Dolan, G. E. Yakubov, M. R. Bonilla, P. Lopez-Sanchezb and J. R. Stokes, Friction, lubrication, and in situ mechanics of poroelastic cellulose hydrogels, *Soft Matter*, 2017, **13**, 3592-3601.
53. R. Michel, L. Poirier, Q. van Poelvoorde, J. Legagneux, M. Manassero and L. Corté, Interfacial fluid transport is a key to hydrogel bioadhesion, *Proc. Natl. Acad. Sci.*, 2019, **116**, 738-743.
54. M. Curatolo, P. Nardinocchi, L. Teresi and D. P. Holmes, Swelling effects on localized adhesion of an elastic ribbon, *Proc. R. Soc. A*, 2019, **475**, 20190067.
55. S. Ekgasit, N. Kaewmanee, P. Jangtawee, C. Thammacharoen and M. Donphoongpri, Elastomeric PDMS Planoconvex Lenses Fabricated by a Confined Sessile Drop Technique, *ACS Appl. Mater. Interfaces*, 2016, **8**, 20474-20482.
56. H. Wen, Characterization of The Surface Properties of Two Pressure Sensitive Adhesives, Masters' Thesis, Johns Hopkins University, 2019.
57. J. D. Glover, C. E. McLaughlin, M. K. McFarland and J. T. Pham, Extracting uncrosslinked material from low modulus sylgard 184 and the effect on mechanical properties, *J. Polym. Sci.*, 2020, **58**, 343-351.
58. I. D. Johnston, D. K. McCluskey, C. K. L. Tan and M. C. Tracey, Mechanical characterization of bulk Sylgard 184 for microfluidics and microengineering, *J. Micromech. Microeng.*, 2014, **24**, 035017.

59. W. Megone, N. Roohpour and J. E. Gautrot, Impact of surface adhesion and sample heterogeneity on the multiscale mechanical characterisation of soft biomaterials, *Sci. Rep.*, 2018, **8**, 6780.
60. †See Electronic Supporting Information (ESI).
61. D. K. Owens and R. C. Wendt, Estimation of Surface Free Energy of Polymers, *J. Appl. Polym. Sci.*, 1969, **13**, 1741–1747.
62. J. N. Lee, C. Park and G. M. Whitesides, Solvent compatibility of poly(dimethylsiloxane)-based microfluidic devices, *Anal. Chem.*, 2003, **75**, 6544-6554.
63. E. Gee, G. Liu, H. Hu and J. Wang, Effect of Varying Chain Length and Content of Poly(dimethylsiloxane) on Dynamic Dewetting Performance of NP-GLIDE Polyurethane Coatings, *Langmuir*, 2018, **34**, 10102-10113.
64. K. Efimenko, W. E. Wallace and J. Genzer, Surface Modification of Sylgard-184 Poly(dimethyl siloxane) Networks by Ultraviolet and Ultraviolet/Ozone Treatment, *J. Colloid Interface Sci.*, 2002, **254**, 306-315.
65. P. Roberts, G. A. Pilkington, Y. M. Wang and J. Frechette, A multifunctional force microscope for soft matter with in situ imaging, *Rev. Sci. Instrum.*, 2018, **89**, 043902.
66. I. K. Lin, K. S. Ou, Y. M. Liao, Y. Liu, K. S. Chen and X. Zhang, Viscoelastic Characterization and Modeling of Polymer Transducers for Biological Applications, *J. Microelectromech. Syst.*, 2009, **18**, 1087-1099.
67. T. Elder, D. Rozairo and A. B. Croll, Origami Inspired Mechanics: Measuring Modulus and Force Recovery with Bent Polymer Films, *Macromolecules*, 2019, **52**, 690-699.
68. D. C. Johnston, Stretched exponential relaxation arising from a continuous sum of exponential decays, *Phys. Rev. B*, 2006, **74**, 184430.
69. D. Cheneler, N. Mehrban and J. Bowen, Spherical indentation analysis of stress relaxation for thin film viscoelastic materials, *Rheol. Acta*, 2013, **52**, 695-706.
70. H. K. Mueller and W. G. Knauss, The Fracture Energy and Some Mechanical Properties of a Polyurethane Elastomer, *Trans. Soc. Rheol.*, 1971, **15**, 217-233.
71. N. Bosnjak, S. Nadimpalli, D. Okumura and S. A. Chester, Experiments and modeling of the viscoelastic behavior of polymeric gels, *J. Mech. Phys. Solids*, 2020, **137**, 103829.
72. H. Z. Ma and J. M. Xu, Mean-Square Radius Fo Gyration of Poly(Dimethylsiloxane) Chain with Side-Groups, *Polym. J.*, 1994, **26**, 779-785.
73. R. Mukherjee and A. Sharma, Instability, self-organization and pattern formation in thin soft films, *Soft Matter*, 2015, **11**, 8717-8740.
74. J. Jopp and R. Yerushalmi-Rozen, Autophobic behavior of polymers at the melt-elastomer interface, *Macromolecules*, 1999, **32**, 7269-7275.
75. P. G. Ferreira, A. Ajdari and L. Leibler, Scaling Law for Entropic Effects at Interfaces between Grafted Layers and Polymer Melts, *Macromolecules*, 1998, **31**, 3994-4003.
76. G. Reiter, A. Sharma, A. Casoli, M. O. David, R. Khanna and P. Auroy, Thin film instability induced by long-range forces, *Langmuir*, 1999, **15**, 2551-2558.
77. C. Redon, F. Brochard-Wyart and F. Rondelez, Dynamics of dewetting, *Phys. Rev. Lett.*, 1991, **66**, 715-718.
78. C. Urata, G. J. Dunderdale, M. W. England and A. Hozumi, Self-lubricating organogels (SLUGs) with exceptional syneresis-induced anti-sticking properties against viscous emulsions and ices, *J. Mater. Chem. A*, 2015, **3**, 12626-12630.
79. D. Maugis, *Contact, Adhesion and Rupture of Elastic Solids*, Springer Berlin, Heidelberg, 1 edn., 2000.

80. A. B. D. Cassie and S. Baxter, Wettability of porous surfaces., *Trans. Faraday Soc.*, 1944, **40**, 0546-0550.
81. P. S. Swain and R. Lipowsky, Contact Angles on Heterogeneous Surfaces: A New Look at Cassie's and Wenzel's Laws, *Langmuir*, 1998, **14**, 6772-6780.
82. É. Degrandi-Contraires, A. Beaumont, F. Restagno, R. Weil, C. Poulard and L. Léger, Cassie-Wenzel–like transition in patterned soft elastomer adhesive contacts, *Europhys. Lett.*, 2013, **101**, 14001.
83. A. R. Mojdehi, D. P. Holmes and D. A. Dillard, Revisiting the generalized scaling law for adhesion: role of compliance and extension to progressive failure, *Soft Matter*, 2017, **13**, 7529-7536.
84. D. Maugis and M. Barquins, Fracture mechanics and the adherence of viscoelastic bodies, *J. Phys. D: Appl. Phys.*, 1978, **11**, 1989-2023.
85. A. N. Gent, G. R. Hamed and W. J. Hung, Adhesion of elastomers: Dwell time effects, *The Journal of Adhesion*, 2003, **79**, 315-325.
86. M. Barquins, Influence of Dwell Time on the Adherence of Elastomers, *The Journal of Adhesion*, 1982, **14**, 63-82.
87. A. N. Gent, Adhesion and Strength of Viscoelastic Solids. Is There a Relationship between Adhesion and Bulk Properties?, *Langmuir*, 1996, **12**, 4492-4496.
88. D. Maugis, Adherence of elastomers: Fracture mechanics aspects, *J. Adhes. Sci. Technol.*, 1987, **1**, 105-134.
89. C. Creton and L. Leibler, How does tack depend on time of contact and contact pressure?, *Journal of Polymer Science Part B-Polymer Physics*, 1996, **34**, 545-554.
90. N. Amouroux and L. Léger, Effect of Dangling Chains on Adhesion Hysteresis of Silicone Elastomers, Probed by JKR Test, *Langmuir*, 2003, **19**, 1396-1401.
91. Y. Sun and G. C. Walker, Viscoelastic Response of Poly(dimethylsiloxane) in the Adhesive Interaction with AFM Tips, *Langmuir*, 2005, **21**, 8694-8702.
92. G. Y. Choi, W. Zurawsky and A. Ulman, Molecular Weight Effects in Adhesion, *Langmuir*, 1999, **15**, 8447-8450.
93. C. Urata, H. Nagashima, B. D. Hatton and A. Hozumi, Transparent Organogel Films Showing Extremely Efficient and Durable Anti-Icing Performance, *ACS Appl. Mater. Interfaces*, 2021, **13**, 28925-28937.
94. J. Delavoipiere, Y. Tran, E. Verneuil and A. Chateauminois, Poroelastic indentation of mechanically confined hydrogel layers, *Soft Matter*, 2016, **12**, 8049-8058.

**Adhesion of fluid infused silicone elastomer to glass**

Anushka Jha[1], Preetika Karnal[1,2], and Joelle Frechette[1,3,*]

*1. Chemical and Biomolecular Engineering Department, Johns Hopkins University, Baltimore, MD 21218, USA 2. Department of Chemical and Biomolecular Engineering, Lehigh University, 124 E Morton St, Building 205, Bethlehem, Pennsylvania 18015, United States 3. Chemical and Biomolecular Engineering Department, University of California, Berkeley, CA 94760, USA*

## Table of Contents

1. Sample preparation

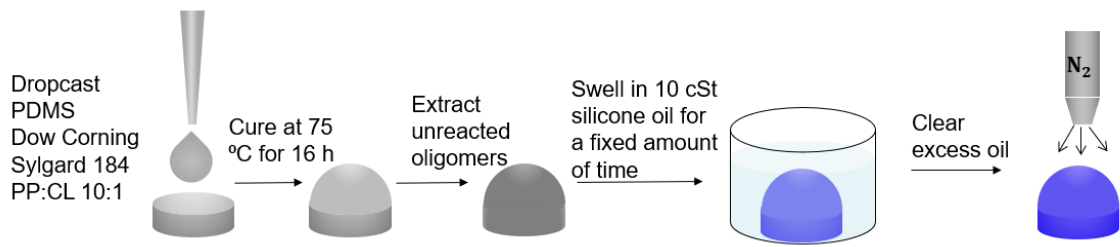

**Fig. S1.** Schematic for dropcasting[1], cure and swelling of PDMS (Sylgard 184).

2. Swelling dynamics of PDMS probe

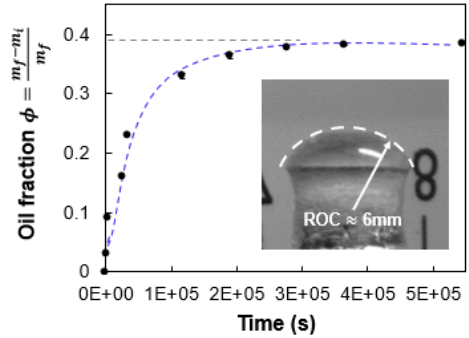

**Fig. S2.** Mass swelling of 10:1 PDMS in 10 cSt silicone oil after hexane extraction.

3. Rheology of dry and swollen PDMS

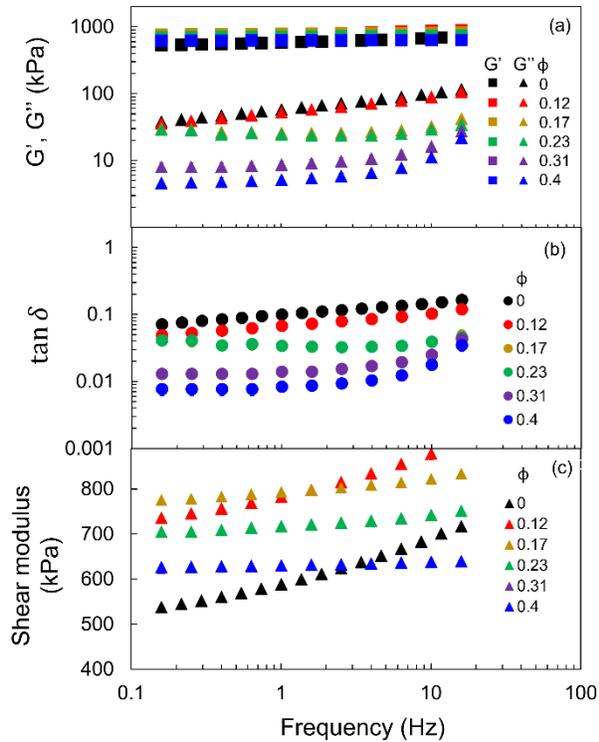

**Fig. S3.** Rheological properties as a function of oil mass fraction of swollen PDMS. (a) Storage modulus (G', squares) and Loss modulus (G'', triangles) (b) $\tan \delta = G''/G'$ (c) Shear modulus $G = \sqrt{G'^2 + G''^2}$ as a function of frequency ($f\ (Hz) = \omega(rad/s)/2\pi$) at different oil fractions.

According to the theory of rubber elasticity (**Eq. S1**), the shear modulus should decrease with increasing oil fraction. This expected decrease is due to reduction in the mass density of crosslinks $\rho/M_C$ as the volume of the elastomer expands as given by:

$$G = \frac{\rho RT}{M_c}. \tag{S1}$$

Here G is the shear modulus of the swollen elastomer calculated from oscillatory rheology (**Fig. S3c**), $\rho$ is the mass density of the elastomer and $M_C$ is the number-average molecular weight of the chain between crosslinks. Instead of a decrease we observe a slight increase in shear modulus from $\phi = 0$ to $\phi = 0.1$, followed by a subsequent decrease with increasing oil fraction. We suspect that the initial deviation from ideal rubber-like behavior is due to the presence of silica fillers in the Sylgard elastomer kit. Upon extraction, some of these silica particles may be removed from the system leaving behind voids that are filled by silicone oil upon initial swelling, without increasing the volume of the swollen sample. Subsequently, as the amount of oil in the bulk increases from $\phi = 0.1$ to $\phi = 0.4$, the overall volume of the elastomer also increases. This results in decreasing density of crosslinks and therefore, decrease in the shear modulus.

4. Connecting swelling and rheology data

The volume swelling ratio, $\Lambda_{swell}$ way can be compared to the equilibrium oil fraction from Flory-Rehner theory.[2, 3] According to Flory-Rehner theory, at equilibrium swelling, the volume fraction of oil $\Lambda_{F-R}$ in the can be expressed using **Eq. (S2)**

$$ln(\Lambda_{F-R}) + (1 - \Lambda_{F-R}) + \chi(1 - \Lambda_{F-R})^2 + \frac{\rho_e V_s}{M_c}(1 - \Lambda_{F-R})^{1/3} = 0 \tag{S2}$$

Where $\chi$ is the Flory Huggins interaction parameter, $V_s$ is the molar volume of the oil. $\rho_e$ is the mass density of the elastomer at maximum swelling and $M_C$ can be obtained from theory of rubber elasticity given in Eq. (S1). Assuming that $\chi \approx 0$ for a chemical similar combination of PDMS elastomer and silicone oil, we obtain $\Lambda_{F-R} = 0.42$ which is similar to the volume oil fraction $\Lambda_{swell}$ obtained from swelling measurements.

5. Choice of frequency for comparison

For a rate of debonding of $v_{deb} = 50$ μm/s, and displacement during detachment: $\delta \sim$ 30-70 μm, we can calculate an effective strain rate (extensional) as $\frac{\delta}{v} \sim 0.6 - 1.4$ Hz which is similar to the chosen rheological frequency(1 Hz). We recognize that the two values may not be analogous due to the different geometry of the experiments but this appears to be a good first approximation.

6. Rate dependence of $G_C$

We assume similar rate dependence for $G_C$ based on the empirical relationship given as[4] $G_C = w(1 + \Phi(a_T v))$ where $w$ is the thermodynamic work of adhesion and $\phi$ is a dimensionless number that depends on the viscoelastic properties of the material and crack speed $v$. It has been seen that $w(\Phi(a_T v)) \propto G'(\omega) h$ [5] where $G'$ is the storage modulus of the system and $h$ is the sample thickness. It can be seen from the plot of $G'$ vs $\phi$ (**Fig S3a**) that $G'$ is similar across samples at different oil fractions. Additionally, we do not see any change in the slope of the $F - \delta$ curve which indicates similar mechanical properties at the given debonding velocity.

7. <u>Moduli from PRI</u>

**Table S1.** Values of effective Young's moduli calculated via **Eq. (4)**.

| $\phi$ | $E_0^*$ | $E_V^*$ | $E_P^*$ |
|---|---|---|---|
| | | | MPa |
| 0 | 3.6 ± 0.4 | 3.0 ± 0.4 | -- |
| 0.1 | 2.8 ± 0.3 | 2.7 ± 0.3 | 2.5 ± 0.3 |
| 0.4 | 3.0 ± 0.5 | 2.9 ± 0.5 | 2.3 ± 0.7 |
| 0.4 (in oil) | 2.6 ± 0.5 | 2.5 ± 0.4 | 2.3 ± 0.4 |

8. <u>Mesh size calculation</u>

Mesh size was calculated by using the following two equations:[6]

$$G = \frac{NkT}{\lambda_0} \quad \text{(Rubber elasticity)}$$

$$\frac{4}{3}\pi \left(\frac{\xi}{2}\right)^3 = \frac{N}{\lambda_0^3} \quad \text{(Chan et al, 2012, Ref 31, main text)}$$

| Symbol | Description | Value |
|---|---|---|
| G | Shear Modulus at swelling equilibrium | $630244 \pm 67192$ Pa |
| N | Number density of crosslinks | mol/m$^3$ |
| k | Boltzmann constant | $1.38 \times 10^{-23}$ J/K |
| T | Temperature | 298.15 K |
| $\lambda_0$ | Equilibrium swelling ratio $= \left(\frac{V_0}{V_i}\right)^{1/3}$ | $1.21 \pm 0.01$ |
| $\xi$ | Mesh size | 2.6 nm |

9. Surface energy determination

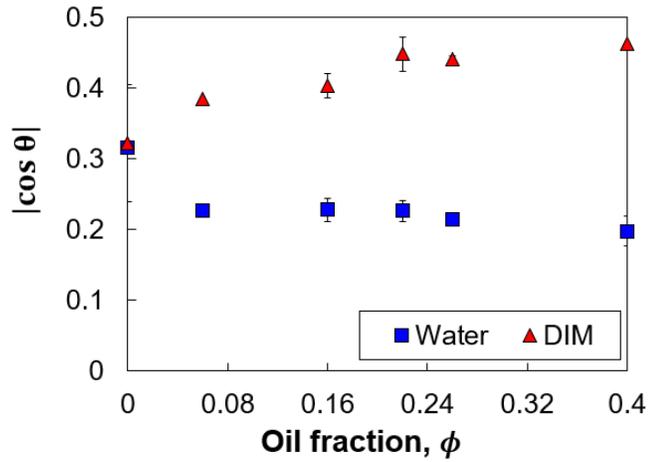

**Fig. S4.** Absolute value of cosine of contact angle, $|cos\ \theta|$ for water (blue squares) and diiodomethane (DIM) (red triangles) on swollen PDMS at different oil fraction. (Numerical values in **Table S3**).

**Table S2.** Reported values of dispersive($\gamma_d$) and polar($\gamma_p$) contributions to surface energy($\gamma$) of water and DIM[7] used to obtain surface energy(SFE) of swollen PDMS using the two liquid method (OWRK)[8] given by **Eq. S3.**

|  | $\gamma$ | $\gamma_p$ | $\gamma_d$ | Dominant component of SFE |
|---|---|---|---|---|
|  | (mJ/m²) | | | |
| **Water (w)** | 72 | 50.4 | 21.6 | Polar |
| **DIM (dim)** | 50 | 2.3 | 47.7 | Dispersive |

$$(1 + \cos \theta_w)\left(\gamma_w^d + \gamma_w^p\right) = 2\left(\sqrt{\gamma_w^d \gamma_s^d} + \sqrt{\gamma_w^p \gamma_s^p}\right)$$
$$(1 + \cos \theta_{dim})\left(\gamma_{dim}^d + \gamma_{dim}^p\right) = 2\left(\sqrt{\gamma_{dim}^d \gamma_s^d} + \sqrt{\gamma_{dim}^p \gamma_s^p}\right) \quad\quad (S3)$$

**Table S3.** Calculation of surface of swollen elastomers from the two-liquid method (OWRK)[7, 8] as a function of oil fractions using **Eq. (S3).**

| ϕ | Contact angle (°) | | cos θ | | $\gamma_s$ (mJ/m²) |
|---|---|---|---|---|---|
|  | DIM | Water | DIM | Water | |
| 0 | 71.2 ± 5.0 | 108.4 ± 0.4 | 0.32 ± 0.08 | −0.32 ± 0.01 | 16.2 ± 1.9 |
| 0.06 | 67.4 ± 0.0 | 103.2 ± 0.0 | 0.38 ± 0.00 | −0.23 ± 0.00 | 18.8 ± 0.0 |
| 0.16 | 66.3 ± 1.1 | 103.2 ± 1.0 | 0.40 ± 0.02 | −0.23 ± 0.02 | 19.0 ± 0.5 |
| 0.22 | 63.4 ± 1.5 | 103.1 ± 0.8 | 0.45 ± 0.02 | −0.23 ± 0.01 | 19.8 ± 0.6 |
| 0.26 | 63.8 ± 0.3 | 102.4 ± 0.6 | 0.44 ± 0.00 | −0.21 ± 0.01 | 19.9 ± 0.2 |
| 0.4 | 62.4 ± 0.0 | 101.4 ± 1.3 | 0.46 ± 0.00 | −0.20 ± 0.02 | 20.7 ± 0.3 |

10. Protocol for adhesion measurements

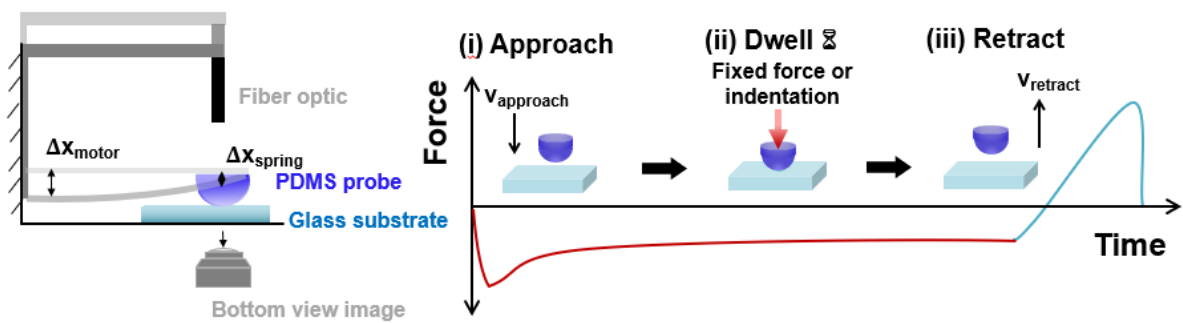

**Fig. S5.** Schematic for PRI and probe tack experiments.

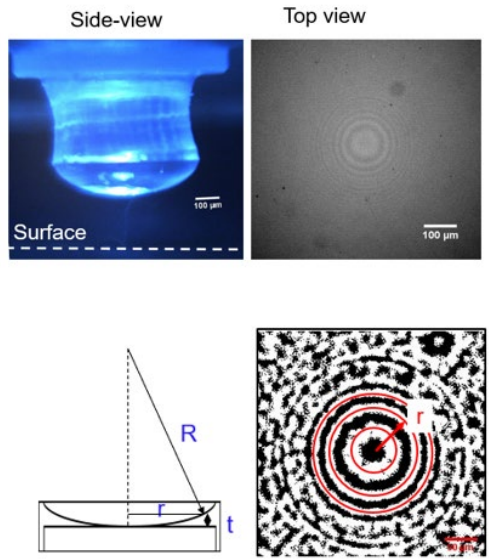

**Fig. S6.** Setting the 'zero' position using side/bottom view imaging.

**Table S4.** Calculation of gap when the lens is in "zero position" and comparison with predictions from thin film interference of visible light.

| Bright ring # | r (µm) | t=r²/2R (nm) | t= (2m+1) λ/4 (nm) | |
|---|---|---|---|---|
| | | | λ=400 nm | λ=700 nm |
| 1 | 33.5 | 102 | 100 | 175 |
| 2 | 66.4 | 400 | 300 | 525 |
| 3 | 89.2 | 724 | 500 | 875 |
| 4 | 108.7 | 1075 | 700 | 1225 |

11. Formation of capillary bridge during debonding of swollen PDMS

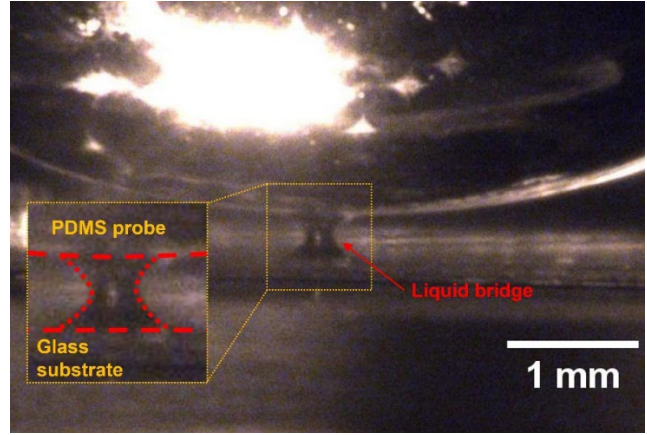

**Fig. S7.** Liquid capillary bridge seen during debonding of fully swollen PDMS from glass.

12. Model for adhesion model that combines capillary and contact forces.

During debonding, the total force acting on the probe, $F_{tot}$ can be assumed to be a sum of the capillary $F_{cap}$ and solid-solid contact force $F_{JKR}$:

$$F_{tot} = F_{cap} + F_{JKR}. \tag{S4}$$

Before the PDMS probe jumps out of contact (**Fig. S8a**), the contribution of capillary force and JKR adhesion to the total force can be given by:

$$F_{cap} = -4\pi\gamma_{oil}R, \tag{S5}$$

$$F_{JKR} = \frac{4}{3}E^*a^3 - 2\sqrt{2\pi w_{adh} E^* a^3}, \tag{S6a}$$

$$\delta = \frac{a^2}{R} - \sqrt{\frac{2\pi w_{adh} a}{E^*}}, \tag{S6b}$$

where $\gamma_{oil} = 20 \; mJ/m^2$ is the surface tension of oil, R is the radius of curvature of the probe, $E^* \approx 2 MPa$ is the effective Young's modulus of the material obtained from rheology, and *a* is contact radius calculated from **Eq. S6b**. (We cannot image the actual solid-solid contact area due to the presence of the oil meniscus.) The work of adhesion $w_{adh}$ in Eq. S7 is calculated by assuming a simple mixing rule given in Eq. S8[9]:

$$w_{adh} = \gamma_{\text{PDMS, air}} + \gamma_{\text{glass, air}} - \gamma_{\text{PDMS, glass}}, \tag{S7}$$

$$w_{adh} \approx 2\sqrt{\gamma_{PDMS,air}\,\gamma_{glass,air}}. \tag{S8}$$

Since, $\gamma_{PDMS} \approx 20$ mJ/m² and $\gamma_{glass} \approx 65$ mJ/m² for pirahna cleaned glass slides, we get $w_{adh} \approx 72$ mJ/m². Solid-solid contact would undergo contact aging with time which is incorporated into our model by varying the work of adhesion using the equation for contact aging:

$$w_{adh}(t) = w_{adh,ref}(t/t_{ref})^n, \tag{S9}$$

and $w_{adh,ref} = 72 \ mJ/m^2$ and $t_{ref} = 100 \ s$ (for simplicity). We assume that dry contact (if any) ages similarly in time as the dry PDMS and use $n = 0.042$ as the power law exponent to obtain an estimate for aging of solid-solid contact. Using $w_{adh}(t)$, we estimate the force as a function of contact time using **Eq. S6a,b** and plotted in **Fig. S9**.

The total force prior to detachment is a sum of the right hand side of **Eq. S5** and **Eq. S6a.** The initial slope of the force-time curve (**Fig. S8c**) is due to the elasticity of the lens contributing to the net compliance of the lens-cantilever system.

After the probe detaches (**Fig. S8b**), the contribution from JKR adhesion vanishes ($F_{JKR} = 0$). The only contribution to the total force is from the capillary bridge[10]:

$$F_{cap} = -\frac{4\pi\gamma_{oil}R\cos\theta}{1 + H/d_{sp\atop pl}} - 2\pi\gamma_{oil}R\sin(\theta + \alpha), \tag{S10}$$

Here θ is the contact angle of oil on glass ≈ 0°, α: embracing angle, H is the height of liquid bridge at the center, $d_{sp\atop pl}$ is the difference between the height of the liquid bridge at the periphery and at the center H. Using an estimate for the volume of the liquid bridge, V, we can calculate both $d_{sp\atop pl}$ and $\alpha$:

$$d_{sp\atop pl} = -H + \sqrt{H^2 + \frac{V}{\pi R}} = \frac{R\alpha^2}{2}. \tag{S11}$$

By fitting the tail of the force time curve (after detachment), we obtain an estimate for the volume of the liquid bridge ~ $0.01\mu L$. We know the values for $\gamma_{oil}$, R, $\theta$ and obtain a theoretical estimate for the capillary force. We see that the liquid meniscus contributes significantly to the overall force, even with the presence of somesolid-solid adhesion at the interface (different lines in **Fig. 8c**).

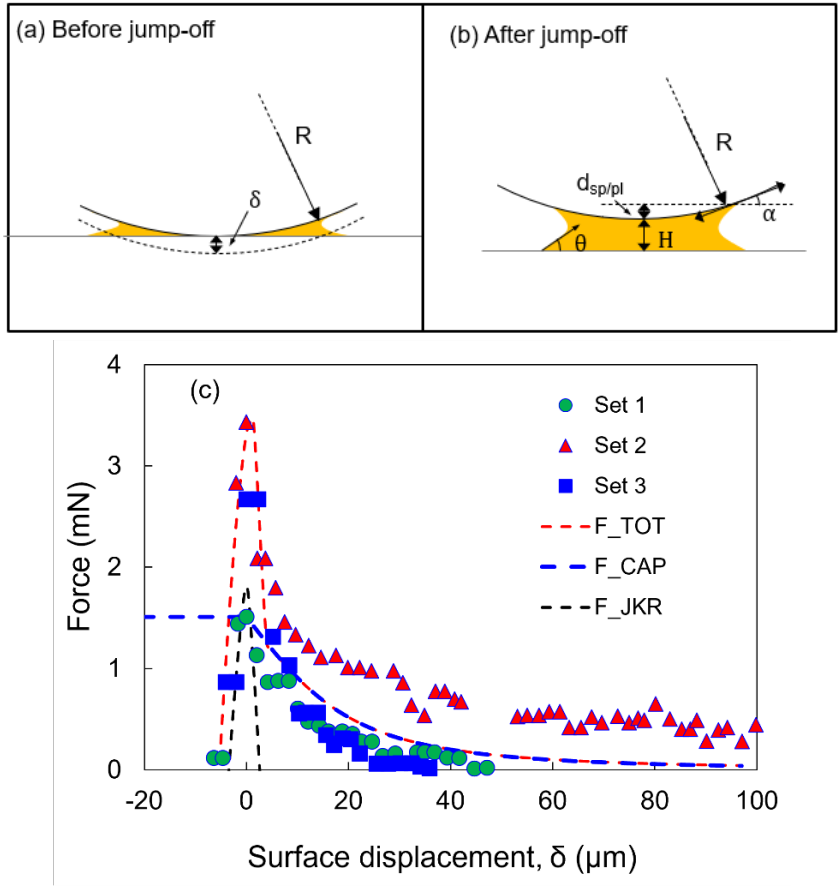

**Fig. S8.** Side view schematic of liquid bridge contact between fully swollen PDMS (sphere) and glass (flat) (a) Before jump off (b) After jump-off (c) Comparing results from probe-tack with capillary/JKR adhesion model (contact time=100s, $w_{adh} = 72 \ mJ/m^2$).

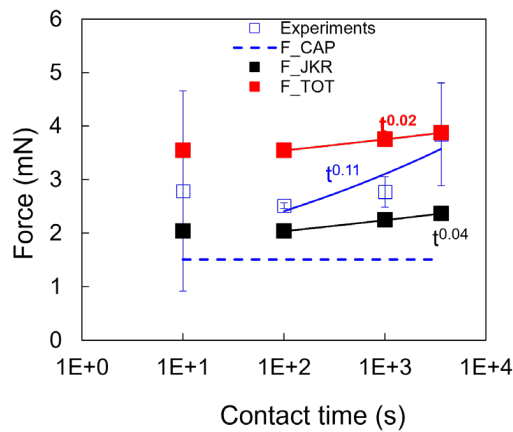

**Fig. S9.** Comparing capillary, JKR pull-off and total pull-off force for different contact times during detachment of fully swollen probe from glass after PRI experiments. Red, blue and black lines represent power law dependence of pull-off force on time.

13. <u>References</u>

1. S. Ekgasit, N. Kaewmanee, P. Jangtawee, C. Thammacharoen and M. Donphoongpri, Elastomeric PDMS Planoconvex Lenses Fabricated by a Confined Sessile Drop Technique, *ACS Appl. Mater. Interfaces*, 2016, **8**, 20474-20482.
2. P. J. Flory, Statistical Mechanics of Swelling of Network Structures, *J. Chem. Phys.*, 1950, **18**, 108-111.
3. P. J. Flory and J. Rehner, Statistical mechanics of cross-linked polymer networks II Swelling, *J. Chem. Phys.*, 1943, **11**, 521-526.
4. D. Maugis and M. Barquins, Fracture mechanics and the adherence of viscoelastic bodies, *J. Phys. D: Appl. Phys.*, 1978, **11**, 1989-2023.
5. C. Creton and M. Ciccotti, Fracture and adhesion of soft materials: a review, *Reports on Progress in Physics*, 2016, **79**, 046601.
6. Y. H. Hu, X. Chen, G. M. Whitesides, J. J. Vlassak and Z. G. Suo, Indentation of polydimethylsiloxane submerged in organic solvents, *J. Mater. Res.*, 2011, **26**, 785-795.
7. K. Efimenko, W. E. Wallace and J. Genzer, Surface Modification of Sylgard-184 Poly(dimethyl siloxane) Networks by Ultraviolet and Ultraviolet/Ozone Treatment, *J. Colloid Interface Sci.*, 2002, **254**, 306-315.
8. D. K. Owens and R. C. Wendt, Estimation of Surface Free Energy of Polymers, *J. Appl. Polym. Sci.*, 1969, **13**, 1741-&.
9. J. N. Israelachvili, in *Intermolecular and Surface Forces (Third Edition)*, ed. J. N. Israelachvili, Academic Press, Boston, 2011, DOI: https://doi.org/10.1016/B978-0-12-391927-4.10017-9, pp. 415-467.
10. Y. I. Rabinovich, M. S. Esayanur and B. M. Moudgil, Capillary forces between two spheres with a fixed volume liquid bridge: Theory and experiment, *Langmuir*, 2005, **21**, 10992-10997.